\documentclass[aps,pra,floatfix,10pt,twocolumn,superscriptaddress,nofootinbib]{revtex4-2}

\usepackage[T1]{fontenc}
\usepackage[utf8]{inputenc}
\usepackage[english]{babel}
\usepackage{CJKutf8} 
\usepackage{microtype}
\usepackage{nicefrac}
\usepackage{soul}        
\setstcolor{red}
\setul{}{1.5pt}
\usepackage{amsmath}
\usepackage{amsfonts}
\usepackage{amssymb}
\usepackage{amstext}
\usepackage{amsthm}
\usepackage{amsbsy}
\usepackage{mathtools}
\usepackage{mathdots}
\usepackage{bm}
\usepackage{dsfont}
\usepackage{braket}
\usepackage{graphicx}
\usepackage{float}
\usepackage{subfloat}      
\usepackage[outdir=./]{epstopdf} 
\usepackage[notransparent]{svg}  
\usepackage{array}
\usepackage{multirow}
\usepackage{makecell}
\usepackage{booktabs}
\newcolumntype{C}{>{$}c<{$}}
\AtBeginDocument{
\heavyrulewidth=.08em
\lightrulewidth=.05em
\cmidrulewidth=.03em
\belowrulesep=.65ex
\belowbottomsep=0pt
\aboverulesep=.4ex
\abovetopsep=0pt
\cmidrulesep=\doublerulesep
\cmidrulekern=.5em
\defaultaddspace=.5em
\tabcolsep=7pt
}
\usepackage{xcolor}
\definecolor{emerald}{rgb}{0.07, 0.53, 0.03}

\usepackage[colorlinks=true,
            citecolor=blue,
            urlcolor=cyan]{hyperref}

\usepackage{nameref}
\usepackage{circuitikz}  
\usepackage{todonotes}  
\usepackage{verbatim}
\makeatletter
\let\latex@addcontentsline\addcontentsline
\renewcommand{\addcontentsline}[3]{}
\makeatother

\newcommand{\fref}[1]{Fig.~\ref{#1}}
\newcommand{\eqsref}[1]{Eq.~\ref{#1}}

\begin{document}

\title{Robust cavity metrology beyond the limits of Pound-Drever-Hall}

\author{Ibukunoluwa Adisa}
\affiliation{Department of Physics, University of Maryland, College Park, MD 20742, USA}
\affiliation{Joint Quantum Institute, NIST/University of Maryland, College Park, Maryland 20742 USA}

\author{Won Chan Lee}
\affiliation{Department of Physics, University of Maryland, College Park, MD 20742, USA}
\affiliation{Joint Quantum Institute, NIST/University of Maryland, College Park, Maryland 20742 USA}

\author{Yanqin Huang}
\affiliation{Department of Physics, University of Maryland, College Park, MD 20742, USA}
\affiliation{Joint Quantum Institute, NIST/University of Maryland, College Park, Maryland 20742 USA}

\author{Nathan Schine}
\affiliation{Department of Physics, University of Maryland, College Park, MD 20742, USA}
\affiliation{Joint Quantum Institute, NIST/University of Maryland, College Park, Maryland 20742 USA}

\author{Kevin C. Cox}
\affiliation{Department of Physics, University of Maryland, College Park, MD 20742, USA}
\affiliation{DEVCOM Army Research Laboratory, 2800 Powder Mill Rd, Adelphi MD 20783, USA}

\author{Alicia J.  Koll\'ar}
\affiliation{Department of Physics, University of Maryland, College Park, MD 20742, USA}
\affiliation{Joint Quantum Institute, NIST/University of Maryland, College Park, Maryland 20742 USA}
\affiliation{Maryland Quantum Materials Center, Department of Physics, University of Maryland, College Park, MD 20742, USA}

\begin{abstract}
Precise measurement of the resonance frequencies of microwave and optical cavities is a foundational capability across quantum technologies.  
One of the highest-performance methods currently available is Pound-Drever-Hall (PDH) spectroscopy, which relies on a three-tone interrogation and eliminates sensitivity to path-length fluctuations.
However, PDH is known to suffer from systematic offsets due to residual amplitude modulation and demodulation-phase errors.
Here we leverage an optimal linear combination of the phases of the three interrogation tones to implement a method for robust cavity metrology which eliminates significant systematic-error sensitivities of PDH.
We experimentally demonstrate robust cavity measurements in both the optical and microwave frequency regime, showing insensitivity to sideband imbalance and demodulation phase, as well as demonstrating single-shot readout of a cavity-coupled superconducting qubit.

\end{abstract}

\maketitle

High-precision measurement of frequency shifts on cavities is a fundamental task across diverse physical platforms, ranging from optical frequency stabilization~\cite{poundMicrowavestabilization,poundDreverHall1983laser, black2001PDHoverview} to superconducting qubit readout~\cite{wallraff2005dispersiveReadout, Blais:CircuitQED}. Near resonance, the phase response of a reflected or transmitted field is linearly sensitive to the frequency difference between the probe field and the cavity, making phase-sensitive measurements a natural tool for cavity spectroscopy. The Pound-Drever-Hall (PDH) technique~\cite{poundMicrowavestabilization,poundDreverHall1983laser, black2001PDHoverview} is one of the most successful methods for this task, utilizing a simple hardware implementation by replacing an absolute phase measurement with a self-referenced carrier-sideband comparison to extract a phase response which is robust to common sources of spurious phase drift. However, imperfections in the modulation and demodulation electronics used to generate the PDH signal are known to produce systematic offsets that limit the performance of high-precision systems~\cite{JunYePDH2024}. Here, we present two measurement observables that generalize PDH and eliminate most of these common systematic error sensitivities, particularly sideband amplitude imbalance and demodulation-phase errors. We show that a combination of standard PDH hardware and an external local oscillator (LO) provides robust spectroscopy signals for both continuous optical-cavity and discrete superconducting-qubit readout.

PDH has enabled some of the most precise optical frequency references, including ultrastable lasers used as phase flywheels between repeated interrogations of atomic ensembles~\cite{robinson2019stabilization, parke2025stabilization}. The performance of these systems is often limited by residual amplitude modulation (RAM)~\cite{whittaker1985RAM, wong1985servoRAM, li2012measurementRAM}, which produces systematic offsets in the PDH discriminator and shifts the locked laser frequency. While active strategies to mitigate RAM exist~\cite{wong1985servoRAM, li2012measurementRAM, hall2015accurateRAM, zhang2014reductionRAM}, they typically rely on extensive calibration involving careful modulation hardware or active compensation. Similarly, superconducting quantum computers require reliable detection of qubit-state-dependent frequency shifts of microwave cavities~\cite{wallraff2005dispersiveReadout, Blais:CircuitQED, quantumengineersguide, practicalGuide_gao2021}. Self-referenced frequency measurements, such as PDH, have the potential to improve the scalability of these systems by reducing hardware synchronization overhead that limits standard microwave architectures~\cite{adisa2026pound}. 

\begin{figure*}[t!]
    \centering
    \includegraphics[width=0.9\linewidth]{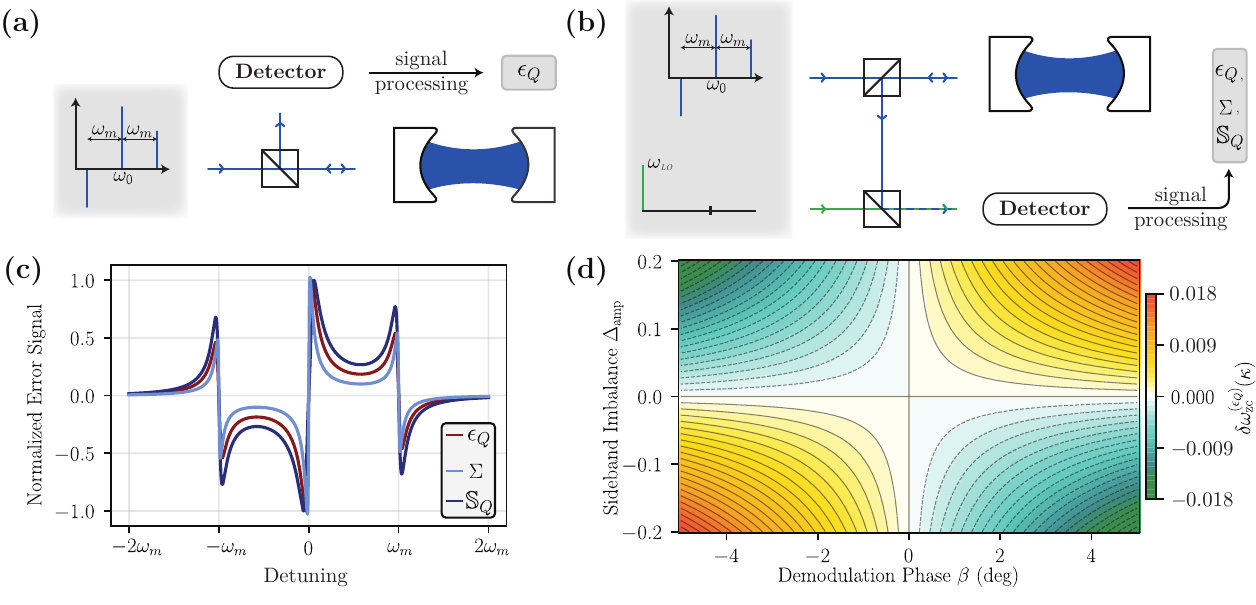}
    \caption{\textit{Three-tone cavity readout and error sensitivity of PDH.}
    (a) Traditional PDH uses a three-tone interrogation scheme in which the carrier at $\omega_0$ and two modulation-induced sidebands at $\omega_{0}\pm\omega_{m}$ interact with a cavity, and square-law detection produces a beatnote signal at $\omega_{m}$; demodulation selects the conventional PDH signal $\epsilon_{Q}$.
    (b) Heterodyne architecture enabling simultaneous extraction of the direct PDH signal and the individual tone quadratures required for reconstructing $\epsilon_{Q}, \Sigma$, and $\mathbb{S}_{Q}$. 
    (c) Comparison of normalized spectroscopy signals ($\epsilon_{Q}, \Sigma$, and $\mathbb{S}_{Q}$) in the ideal limit, demonstrating that all three quantities provide similar dispersive discriminators with a useful zero-crossing and linear slope near resonance.
    (d) Shift $\delta\omega_{\mathrm{zc}}^{(\epsilon_{Q})}$ of the zero-crossing of $\epsilon_{Q}$ versus sideband imbalance $\Delta_{\mathrm{amp}}$ and demodulation phase error $\beta$ demonstrates the susceptibility of conventional PDH to systematic errors.
    }
    \label{fig:schematics_error_signals}
\end{figure*}

The conventional PDH technique uses a three-tone field, typically generated by phase modulation at frequency $\omega_{m}$, to interrogate the cavity, as shown schematically in~\fref{fig:schematics_error_signals}a. A carrier at frequency $\omega_{0}$ and the two sidebands at $\omega_{\pm} = \omega_{0} \pm \omega_{m}$ interact with the cavity, and square-law (intensity) detection produces a beat at $\omega_{m}$ between the carrier and sidebands. Demodulation of one quadrature of the beat pattern is used to select the conventional PDH signal
\begin{equation} \label{eqs:pdh_q}
    \epsilon_{Q} = E_{0}E_{+}\sin{(\phi_{0} - \phi_{+})} - E_{0}E_{-}\sin{(\phi_{0} - \phi_{-})},
\end{equation}
where $E_{j}$ and $\phi_{j}$ are the detected amplitude and phase of each tone $j \in \{-,0,+\}$, for the lower sideband ($-$), carrier ($0$), and the upper sideband ($+$). In the ideal limit (perfect phase modulation with equal sidebands), $\epsilon_{Q}$ is a dispersive spectroscopy signal with a linear slope and zero-crossing centered at resonance, as shown in \fref{fig:schematics_error_signals}c. Much of the success of PDH comes from the fact that $\epsilon_{Q}$ depends only on carrier-sideband phase difference, so that phase shifts common to all three tones cancel out. Interference between the sidebands also makes the signal vanish on resonance independent of the overall optical power. However, imperfections in the modulation and detection setup can distort the measured signal $\epsilon_{Q}$, systematically shifting the zero-crossing away from the true cavity resonance, as shown in \fref{fig:schematics_error_signals}d. The method presented here, based on the scissors phase $\Sigma = 2\phi_{0} - (\phi_{-} + \phi_{+})$, introduced in Ref.~\cite{adisa2026pound} and detected using a three-frequency heterodyne setup (\fref{fig:schematics_error_signals}b), is intrinsically robust to many of the systematic vulnerabilities of PDH while preserving its useful discriminator structure (\fref{fig:schematics_error_signals}c). 

Here, we first formulate conventional PDH and the scissors-phase observables within a common three-tone framework and identify their sensitivities to the relevant error modes, summarized in Table~\ref{table:systematics}. We then benchmark their robustness experimentally in two distinct physical regimes. In an optical experiment, we demonstrate continuous frequency discrimination, where a shifted zero-crossing directly represents a lock-point error. In a microwave experiment, we show discrete qubit-state discrimination, where the relevant metric is the resolvability of the measured ground- and excited-state distributions under common RF errors. 

\textit{Systematic Error Processes}.--- To understand the systematic vulnerabilities of PDH, it is helpful to consider all 3 tones at once and consider the possible types of correlations between the amplitudes and phases. We write a general set of phases as a linear combination of orthogonal, collective phase modes
\begin{align}
    \nonumber
    (\phi_{-}, \phi_{0}, \phi_{+}) &= \mathcal{A} (1,1,1) \\
    & \quad + \mathcal{B} (-1, 0, 1) + \tfrac{\Sigma}{6} (-1,2,-1).
\end{align}
Each phase mode represents a distinct physical process in the hardware: $\mathcal{A}$ is a common-mode phase that effects all three tones equally,  
$\mathcal{B}$ is a differential-mode phase that advances one sideband while delaying the other, and
$\Sigma$ is the scissors phase, which takes the value of $\pi$ for an ideal phase-modulated signal. (See Section~\ref{sec:modulation_imperfections} of the Supplemental Material for details of the relationship between the phase modes, amplitude/phase modulation, and RAM.)
Next we consider possible deviations which change the phases to
\begin{align}
    \nonumber
    (\phi_-', \phi_0', \phi_+') = &(\phi_-, \phi_0, \phi_+) + \alpha (1,1,1)  \\
    \nonumber
    &+ \beta (-1,0,1) + \sigma (-1,2,-1).
\end{align}
Common-mode deviations $(\alpha)$ are generally the largest, arising from path-length fluctuations or drifts in shared references. 
Differential-mode deviations $(\beta)$ are typically smaller, and usually arise from (de)modulation-phase miscalibrations. 
Scissors-mode deviations $(\sigma)$ from sources other than the cavity are generally very small, requiring the carrier phase to shift relative to the average phase of the two sidebands. Such deviations require quadratic variation of the hardware phase response across the modulation bandwidth. Because the collective phase modes are orthogonal, measuring $\Sigma$ detects the desired cavity-induced phase shift while being completely insensitive to $\mathcal{A}$ and $\mathcal{B}$, which are prone to technical error.

From the point of view of collective phase modes, the conventional PDH signal $\epsilon_{Q}$ can be viewed as an approximation to $\Sigma$: 
in the limit of perfect demodulation phase $\mathcal{B} = 0$, balanced sidebands, and weak coupling,
$\epsilon_{Q} \propto E_{0}E_{+}\Sigma$ (see Section~\ref{sec:weak_coupling} in the Supplemental Material).
Away from this limit, the amplitude weighting in \eqsref{eqs:pdh_q} and imperfect demodulation cause the measured PDH signal to deviate from the underlying scissors phase. In particular, a differential-mode phase error rotates the measured PDH signal, while sideband imbalance ($E_{+} \neq E_{-}$) distorts the cancellation between the beatnotes produced by the two sidebands. When both effects are present, the measured PDH signal is no longer proportional to the scissors phase. This deviation appears experimentally as RAM and a systematic offset of the PDH zero-crossing $\omega_{zc}^{(\epsilon_{Q})}$.


\begin{table}[t!]
\centering
\caption{\textit{Sensitivity of cavity-readout observables to systematic errors.}
$\checkmark$ denotes rejection of the error, 
$\large\mathrm{X}$ denotes sensitivity,
n.a. denotes an inapplicable error mode, and 
M denotes the primary measurement observable.
-- Single-tone heterodyne relies solely on the global phase ($\mathcal{A}$) which is strongly affected by path-length fluctuations and synchronization errors. PDH is robust to these errors, but will develop systematic offsets in the presence of amplitude imbalance $(\Delta_{\rm amp})$ and modulation/demodulation phase errors $(\mathcal{B})$. The scissors observables ($\Sigma$ and $\mathbb{S}_Q$) are robust to all these error channels.
\label{table:systematics}
}
\begin{tabular*}{\columnwidth}{@{\extracolsep{\fill}}l|cccc@{}}
\hline\hline
   & Global & Modulation &  $\Sigma$ & Sideband \\
   & Phase & Phase &   & Imbalance \\
\hline
Heterodyne         & M      & n.a.            & n.a. & n.a. \\
PDH $\epsilon_Q$   & $\checkmark$  & $\large\mathrm{X}$      & M  & $\large\mathrm{X}$ \\
$\Sigma$           & $\checkmark$  & $\checkmark$  & M  & $\checkmark$ \\
$\mathbb{S}_Q$     & $\checkmark$  & $\checkmark$  & M  & $\checkmark$ \\
\hline\hline
\end{tabular*}
\end{table}

In general, in the presence of a fractional sideband imbalance $\Delta_{\mathrm{amp}}$ such that $E_{\pm}=E_{s}(1 \pm \Delta_{\mathrm{amp}}/2)$, the changes in the zero-crossing of the PDH signal can be written as
\begin{equation} \label{eqs:pdh_zerocrossing}
    \delta\omega_{\text{zc}}^{(\epsilon_{Q})} \propto -\kappa \left( 3\sigma  - \frac{\beta\Delta_{\mathrm{amp}}}{2} \right),
\end{equation}
where $\kappa$ is the linewidth of the cavity. 
The term $-3\sigma$ is the desired shift of the cavity-response coordinate, which will move the lock point if the underlying cavity frequency has moved. 
The other term proportional to $\beta \Delta_{\mathrm{amp}}$ is a false shift caused by sideband imbalance leaking into the PDH signal through a differential phase error. The effect of their product is the zero-crossing displacement captured by the contour lines in \fref{fig:schematics_error_signals}d.  

To eliminate the distortion that causes the zero-crossing shifts, we directly reconstruct the phase that ideal PDH only approximates: $\Sigma$. The heterodyne architecture in \fref{fig:schematics_error_signals}b makes this possible by adding an LO which enables heterodyne detection of the reflected response of all three tones individually, rather than only as a single analog beatnote. Measurement of the individual phases allows a direct computation of $\Sigma$. Because $\Sigma$ structurally rejects $\alpha$ and $\beta$ regardless of amplitude imbalance, its zero-crossing remains fundamentally locked to the cavity resonance, providing a superior robust observable for high-precision measurements. Mathematically, the shift in the zero-crossing of $\Sigma$ is 
\begin{equation} \label{eqs:sigma_zerocrossing}
    \delta\omega_{\text{zc}}^{(\Sigma)} \propto -3 \sigma \kappa ,
\end{equation}
with no dependence on either $\beta$ and $\Delta_{\mathrm{amp}}$ (see Section~\ref{sec:zero_crossing} of the Supplemental Information). However, the explicit phase extraction needed to compute $\Sigma$ introduces its own computational cost as arctangent operations to compute $\phi_j$ suffer from phase wrapping errors, behave poorly on low signal-to-noise signals, and can be slow for real-time feedback and qubit readout. 

\begin{figure}[t!]
    \centering
    \includegraphics[width=0.48\textwidth]{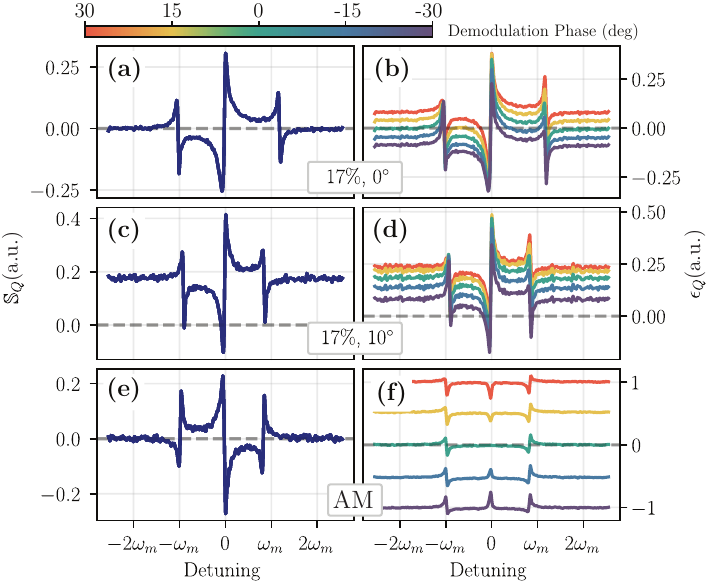}
    \caption{\textit{Optical spectroscopy signals under induced hardware imperfections}.
    $\mathbb{S}_{Q}$ (a) and $\epsilon_{Q}$ (b) versus cavity detuning, with $17\%$ sideband power imbalance, $0^{\circ}$ scissors-phase offset, and varying demodulation phase. Line color indicates applied demodulation phase.
    $\mathbb{S}_{Q}$ (c) and $\epsilon_{Q}$ (d) versus cavity detuning, with $17\%$ sideband power imbalance, $10^{\circ}$ scissors-phase offset, and varying demodulation phase.
    (e) For AM $\mathbb{S}_{Q}$ remains a stable discriminator for locking unlike $\epsilon_{Q}$ (f). All traces in each panel are normalized by the average off-resonant value of $E_{+}E_{-}E_{0}^{2}$ (for $\mathbb{S}_{Q}$) or $E_{0}(E_{-}+E_{+})/2$ (for $\epsilon_Q$). Additional configurations are shown in \fref{fig:supple_optical_imbalances} of the Supplemental Material.
    }
    \label{fig:error_signals_demod}
\end{figure}

To bypass these issues, we introduce an experimentally efficient scissors-phase observable $\mathbb{S}$, whose primary quadrature of interest is 
\begin{equation}
    \mathbb{S}_{Q} = E_{-}E_{+}E_{0}^{2} \sin{\Sigma},
\end{equation}
which retains the same dispersive discriminator behavior as $\Sigma$ and $\epsilon_{Q}$ (see \fref{fig:schematics_error_signals}c, and see Section~\ref{sec:sq_arctangent_free} of the Supplemental Material for full derivation). Because $\mathbb{S}_{Q}$ is a direct function of $\Sigma$, it also inherits rejection of common-mode and differential-mode phase errors. The practical advantage of $\mathbb{S}_{Q}$ is that it avoids any arctangent computation and can be computed directly from the measured IQ components without evaluating $\Sigma$ itself. Let $\mathbf{v}_{j} = (I_{j}, Q_{j}, 0)$ be the measured IQ heterodyne vector of tone $j$,
\begin{align}
    \mathbb{S}_{Q} &= [\mathbf{v}_{+} \times \mathbf{v}_{0}]_{\mathrm{z}} [\mathbf{v}_{-} \cdot \mathbf{v}_{0}] + [\mathbf{v}_{-} \times \mathbf{v}_{0}]_{\mathrm{z}} [\mathbf{v}_{+} \cdot \mathbf{v}_{0}]  \\
    \nonumber
    & =  (Q_0I_+ - I_0Q_+)(Q_0Q_- + I_0I_-) \\
    & \quad + (Q_0I_- - I_0Q_-)(Q_0Q_+ + I_0I_+).     
\end{align}

\begin{figure*}[t!]
    \centering
    \includegraphics[width=0.85\linewidth]{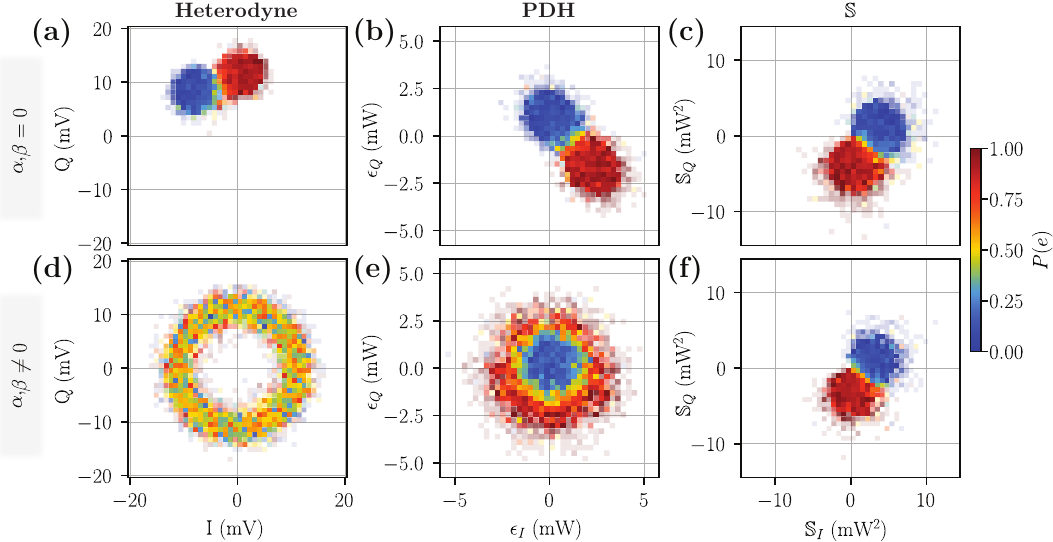}
    \caption{\textit{Superconducting-qubit readout under controlled phase errors.}
    Single-shot IQ distributions for conventional carrier-heterodyne (a), PDH (b), and $\mathbb{S}$ (c) under ideal phase-locked conditions ($\alpha, \beta = 0$). All three observables yield well-separated regions corresponding to $\ket{g}$ (blue) and $\ket{e}$ (red). 
    The color scale indicates the probability $P(e)$ of events corresponding to preparation in $\ket{e}$. 
    Bins containing fewer than 10 counts displayed with proportionally reduced opacity (see Section~\ref{sec:extended_microwave_data} in Supplemental Material).
    (d-f) Corresponding readout distributions under combined common-mode and differential-mode errors ($\alpha, \beta \neq 0$). Heterodyne results in overlapping and indistinguishable distributions of $\ket{g}$ and $\ket{e}$. PDH is partially degraded by random rotation between $\epsilon_I$ and $\epsilon_Q$, creating a larger yellow region of overlap.
    $\mathbb{S}$ remains unaffected. All datasets include $5000$ shots with the qubit prepared in $\ket{g}$ and $5000$ shots with the qubit prepared in $\ket{e}$.
    }
    \label{fig:microwave_IQ_main}
\end{figure*}

Unlike the scissors phase $\Sigma$, $\mathbb{S}_{Q}$ uses only simple addition and multiplication of the raw heterodyne components, making it purely single-valued and efficient for real-time digital processing. Similarly to $\Sigma$, the shift in the zero-crossing of $\mathbb{S}_{Q}$, $\delta\omega_{\text{zc}}^{(\mathbb{S}_{Q})} \propto -3 \sigma \kappa $. However, since $\mathbb{S}_{Q}\propto E_{-}E_{+}E_{0}^{2}$, sideband imbalance rescales the signal, but does not change the value of $\Sigma$ at which $\mathbb{S}_{Q} = 0$. Thus $\mathbb{S}_{Q}$ provides a computationally efficient measure of the scissors phase while preserving its robust zero crossing. The ideal forms of these three observables ($\epsilon_{Q}$, $\Sigma$, and $\mathbb{S}_{Q}$) are as shown in \fref{fig:schematics_error_signals}c and a comparison of the systematic error susceptibilities is shown in Table~\ref{table:systematics}.

\textit{Optical implementation: continuous frequency discrimination}.---
Accurate frequency stabilization between a laser and an optical cavity requires an error signal whose zero crossing is tied to cavity response. Any offset in the error signal is converted by feedback into a finite detuning of the laser from the cavity. Here, we generate $\epsilon_{Q}$ and $\mathbb{S}_{Q}$ for an optical cavity and demonstrate the superior systematic-error immunity of $\mathbb{S}_{Q}$. 


We implement the heterodyne measurement scheme, described in \fref{fig:schematics_error_signals}b, which allows $\epsilon_{Q}$ and $\mathbb{S}_{Q}$ of an optical cavity to be computed from the same measurement record. We then deliberately introduce nonidealities in the measurement, including sideband amplitude imbalance, demodulation phase errors, and a controlled scissors-phase offset. With a sideband power imbalance and no imposed scissors-mode shift, $\mathbb{S}_{Q}$ remains centered at zero as the demodulation phase is varied (see \fref{fig:error_signals_demod}a showing an example with $17\%$ power imbalance). On the other hand, $\epsilon_{Q}$ acquires a demodulation-phase-dependent offset (\fref{fig:error_signals_demod}b) because the sideband imbalance allows demodulation-phase errors to enter the PDH signal, as shown in \eqsref{eqs:pdh_zerocrossing}.

The immunity of $\mathbb{S}_{Q}$ to demodulation phase and amplitude errors demonstrates its advantage as a robust frequency discriminator. This same distinction persists when the desired scissors phase is deliberately shifted. When subjected to sideband power imbalance and scissors-phase offset, $\mathbb{S}_{Q}$ shifts uniformly and remains insensitive to the demodulation phase (see \fref{fig:error_signals_demod}c showing an example with a $17\%$ sideband power imbalance and $10^{\circ}$ scissors-phase offset). In contrast, the conventional PDH signal $\epsilon_{Q}$ continues to show demodulation dependent shifts and distortions (\fref{fig:error_signals_demod}d). Thus, $\mathbb{S}_{Q}$ preserves sensitivity to the cavity response while rejecting the technical offsets that move the PDH zero-crossing. 

Futhermore, we demonstrate the robustness of $\mathbb{S}_{Q}$ when the modulation convention itself is changed. When the optical cavity is driven purely by amplitude modulation (AM) instead of the standard phase modulation, the PDH signal $\epsilon_{Q}$ loses its dispersive PDH-like characteristics entirely, with no zero-crossing on resonance, and becomes unsuitable for locking or dispersive readout (\fref{fig:error_signals_demod}f). However, $\mathbb{S}_{Q}$ retains its dispersive discriminator because the change to AM shifts the scissors phase by $\pi$, which only simply reverses the sign of $\mathbb{S}_{Q}$ (\fref{fig:error_signals_demod}e). 



\textit{Microwave implementation: discrete state discrimination}.---
Dispersive readout of a superconducting qubit requires distinguishing discrete, state-dependent frequency shifts of a microwave resonator in an environment with significant noise~\cite{wallraff2005dispersiveReadout}. While standard synthetic PDH readout suppresses the common-mode phase drifts that plague absolute heterodyne measurements, it remains vulnerable to radio-frequency (RF) timing errors~\cite{adisa2026pound}. Here, we show that the scissors observables eliminate this sensitivity entirely.

Using the microwave analog of \fref{fig:schematics_error_signals}b, we compare three cases $\mathbb{S}$, PDH, and single-tone heterodyne using the carrier tone only, which is the standard technique for superconducting-qubit readout~\cite{wallraff2005dispersiveReadout, quantumengineersguide}.
(see Section~\ref{sec:microwave_setup} of the Supplemental Material for the experimental setup).
All observables are computed from the same single-shot measurements, ensuring that any difference in state discrimination arises from the observable used, and not from a change in the pulse sequence, resonator response, or qubit preparation. 
For each configuration, we collect $5000$ single-shot measurements with the qubit prepared in $\ket{g}$ and $5000$ with the qubit prepared in $\ket{e}$. 
IQ-plane histograms of the measurement outcomes are shown in Fig.~\ref{fig:microwave_IQ_main}.
The color of each pixel indicates probability that events in that pixel correspond to preparation in the $\ket{e}$ state.
When all microwave components are locked and synchronized (Fig.~\ref{fig:microwave_IQ_main}(a-c)) all three observables provide clear distinct regions in the IQ plane corresponding to $\ket{g}$ (blue) and $\ket{e}$ (red).
Next, we inject common-mode and differential-mode errors by unlocking the generators and adding time delays to the modulation and data acquisition. Under these conditions, heterodyne readout no longer distinguishes the two  qubit states. 
PDH readout is insensitive to common-mode errors, but differential-mode errors cause rotations between $\epsilon_Q$ and $\epsilon_I$, reducing the distinguishability between $\ket{g}$ and $\ket{e}$.
(See \fref{fig:supple_microwave_IQs} in the Supplemental Material for measurements with common-mode errors only.) 
$\mathbb{S}$, on the other hand, is unaffected and still provides clear distinguishability between the two qubit states.
See Section~\ref{sec:extended_microwave_data} of the Supplemental Material for determination of resolvability and assignment fidelity for
$\epsilon$, $\Sigma$, and $\mathbb{S}$.

\textit{Outlook}.--- Conventional PDH remains a state-of-the-art self-referenced measurement tool across diverse physical platforms because it avoids the absolute carrier-phase sensitivity of heterodyne, relying instead on carrier-sideband phase differences, and suppressing the usually dominant common-mode phase drift. 
However, the measured PDH signal may still acquire erroneous offsets from residual amplitude modulation and demodulation-phase errors. The scissors phase $\Sigma$ and its computationally efficient alternative $\mathbb{S}_{Q}$ provide robust alternative observables that eliminate systematic errors associated with sideband-amplitude imbalance and demodulation phase.
Cavity metrology using $\Sigma$ or $\mathbb{S}_{Q}$ relaxes phase-noise and synchronization requirements,
thereby reducing the experimental hardware and calibration overhead needed for accurate cavity metrology, and may improve fundamental accuracy limits. 
In a future where quantum technologies become commonplace, scissors-phase metrology can simplify the hardware required for precision metrology, scalable quantum information processing, and portable quantum devices.

\begin{acknowledgements}
This project was supported by Army-Maryland Partnership for Quantum Electrodynamics (ARL Grant No. W911NF-24-2-0107).
YH was supported by ARL (Grant No. W911NF-24-2-0107).
IA, WCL, and AK received support from ARL (Grant Nos. W911NF-19-2-0181 and W911NF-17-S-0003), NSF (Grant No. PHY2047732), AFOSR (Grant No. FA9550-21-1-0129), NSF QLCI (Grant No. OMA-2120757), the University of Maryland, and the Maryland Quantum Materials Center.
We thank Jun Ye, Jim Phillips, and Trey Porto for helpful comments and discussions. We thank Mariia Kharchenko and Tara Taneja for contributions to the laser system.
\end{acknowledgements}







\bibliographystyle{apsrev4-2}
\bibliography{refs.bib}

\begin{thebibliography}{18}%
\makeatletter
\providecommand \@ifxundefined [1]{%
 \@ifx{#1\undefined}
}%
\providecommand \@ifnum [1]{%
 \ifnum #1\expandafter \@firstoftwo
 \else \expandafter \@secondoftwo
 \fi
}%
\providecommand \@ifx [1]{%
 \ifx #1\expandafter \@firstoftwo
 \else \expandafter \@secondoftwo
 \fi
}%
\providecommand \natexlab [1]{#1}%
\providecommand \enquote  [1]{``#1''}%
\providecommand \bibnamefont  [1]{#1}%
\providecommand \bibfnamefont [1]{#1}%
\providecommand \citenamefont [1]{#1}%
\providecommand \href@noop [0]{\@secondoftwo}%
\providecommand \href [0]{\begingroup \@sanitize@url \@href}%
\providecommand \@href[1]{\@@startlink{#1}\@@href}%
\providecommand \@@href[1]{\endgroup#1\@@endlink}%
\providecommand \@sanitize@url [0]{\catcode `\\12\catcode `\$12\catcode `\&12\catcode `\#12\catcode `\^12\catcode `\_12\catcode `\%12\relax}%
\providecommand \@@startlink[1]{}%
\providecommand \@@endlink[0]{}%
\providecommand \url  [0]{\begingroup\@sanitize@url \@url }%
\providecommand \@url [1]{\endgroup\@href {#1}{\urlprefix }}%
\providecommand \urlprefix  [0]{URL }%
\providecommand \Eprint [0]{\href }%
\providecommand \doibase [0]{http://dx.doi.org/}%
\providecommand \selectlanguage [0]{\@gobble}%
\providecommand \bibinfo  [0]{\@secondoftwo}%
\providecommand \bibfield  [0]{\@secondoftwo}%
\providecommand \translation [1]{[#1]}%
\providecommand \BibitemOpen [0]{}%
\providecommand \bibitemStop [0]{}%
\providecommand \bibitemNoStop [0]{.\EOS\space}%
\providecommand \EOS [0]{\spacefactor3000\relax}%
\providecommand \BibitemShut  [1]{\csname bibitem#1\endcsname}%
\let\auto@bib@innerbib\@empty
\bibitem [{\citenamefont {Pound}(1946)}]{poundMicrowavestabilization}%
  \BibitemOpen
  \bibfield  {author} {\bibinfo {author} {\bibfnamefont {R.~V.}\ \bibnamefont {Pound}},\ }\bibfield  {title} {\enquote {\bibinfo {title} {Electronic frequency stabilization of microwave oscillators},}\ }\href@noop {} {\bibfield  {journal} {\bibinfo  {journal} {Review of Scientific Instruments}\ }\textbf {\bibinfo {volume} {17}},\ \bibinfo {pages} {490} (\bibinfo {year} {1946})}\BibitemShut {NoStop}%
\bibitem [{\citenamefont {Drever}\ \emph {et~al.}(1983)\citenamefont {Drever}, \citenamefont {Hall}, \citenamefont {Kowalski}, \citenamefont {Hough}, \citenamefont {Ford}, \citenamefont {Munley},\ and\ \citenamefont {Ward}}]{poundDreverHall1983laser}%
  \BibitemOpen
  \bibfield  {author} {\bibinfo {author} {\bibfnamefont {R.~W.}\ \bibnamefont {Drever}}, \bibinfo {author} {\bibfnamefont {J.~L.}\ \bibnamefont {Hall}}, \bibinfo {author} {\bibfnamefont {F.~V.}\ \bibnamefont {Kowalski}}, \bibinfo {author} {\bibfnamefont {J.}~\bibnamefont {Hough}}, \bibinfo {author} {\bibfnamefont {G.}~\bibnamefont {Ford}}, \bibinfo {author} {\bibfnamefont {A.}~\bibnamefont {Munley}}, \ and\ \bibinfo {author} {\bibfnamefont {H.}~\bibnamefont {Ward}},\ }\bibfield  {title} {\enquote {\bibinfo {title} {Laser phase and frequency stabilization using an optical resonator},}\ }\href@noop {} {\bibfield  {journal} {\bibinfo  {journal} {Applied Physics B}\ }\textbf {\bibinfo {volume} {31}},\ \bibinfo {pages} {97} (\bibinfo {year} {1983})}\BibitemShut {NoStop}%
\bibitem [{\citenamefont {Black}(2001)}]{black2001PDHoverview}%
  \BibitemOpen
  \bibfield  {author} {\bibinfo {author} {\bibfnamefont {E.~D.}\ \bibnamefont {Black}},\ }\bibfield  {title} {\enquote {\bibinfo {title} {An introduction to pound--drever--hall laser frequency stabilization},}\ }\href@noop {} {\bibfield  {journal} {\bibinfo  {journal} {American journal of physics}\ }\textbf {\bibinfo {volume} {69}},\ \bibinfo {pages} {79} (\bibinfo {year} {2001})}\BibitemShut {NoStop}%
\bibitem [{\citenamefont {Wallraff}\ \emph {et~al.}(2005)\citenamefont {Wallraff}, \citenamefont {Schuster}, \citenamefont {Blais}, \citenamefont {Frunzio}, \citenamefont {Majer}, \citenamefont {Devoret}, \citenamefont {Girvin},\ and\ \citenamefont {Schoelkopf}}]{wallraff2005dispersiveReadout}%
  \BibitemOpen
  \bibfield  {author} {\bibinfo {author} {\bibfnamefont {A.}~\bibnamefont {Wallraff}}, \bibinfo {author} {\bibfnamefont {D.~I.}\ \bibnamefont {Schuster}}, \bibinfo {author} {\bibfnamefont {A.}~\bibnamefont {Blais}}, \bibinfo {author} {\bibfnamefont {L.}~\bibnamefont {Frunzio}}, \bibinfo {author} {\bibfnamefont {J.}~\bibnamefont {Majer}}, \bibinfo {author} {\bibfnamefont {M.~H.}\ \bibnamefont {Devoret}}, \bibinfo {author} {\bibfnamefont {S.~M.}\ \bibnamefont {Girvin}}, \ and\ \bibinfo {author} {\bibfnamefont {R.~J.}\ \bibnamefont {Schoelkopf}},\ }\bibfield  {title} {\enquote {\bibinfo {title} {Approaching unit visibility for control of a superconducting qubit with dispersive readout},}\ }\href@noop {} {\bibfield  {journal} {\bibinfo  {journal} {Physical review letters}\ }\textbf {\bibinfo {volume} {95}},\ \bibinfo {pages} {060501} (\bibinfo {year} {2005})}\BibitemShut {NoStop}%
\bibitem [{\citenamefont {Blais}\ \emph {et~al.}(2021)\citenamefont {Blais}, \citenamefont {Grimsmo}, \citenamefont {Girvin},\ and\ \citenamefont {Wallraff}}]{Blais:CircuitQED}%
  \BibitemOpen
  \bibfield  {author} {\bibinfo {author} {\bibfnamefont {A.}~\bibnamefont {Blais}}, \bibinfo {author} {\bibfnamefont {A.~L.}\ \bibnamefont {Grimsmo}}, \bibinfo {author} {\bibfnamefont {S.~M.}\ \bibnamefont {Girvin}}, \ and\ \bibinfo {author} {\bibfnamefont {A.}~\bibnamefont {Wallraff}},\ }\bibfield  {title} {\enquote {\bibinfo {title} {Circuit quantum electrodynamics},}\ }\href {\doibase 10.1103/RevModPhys.93.025005} {\bibfield  {journal} {\bibinfo  {journal} {Rev. Mod. Phys.}\ }\textbf {\bibinfo {volume} {93}},\ \bibinfo {pages} {025005} (\bibinfo {year} {2021})}\BibitemShut {NoStop}%
\bibitem [{\citenamefont {Kedar}\ \emph {et~al.}(2024)\citenamefont {Kedar}, \citenamefont {Yao}, \citenamefont {Ryger}, \citenamefont {Hall},\ and\ \citenamefont {Ye}}]{JunYePDH2024}%
  \BibitemOpen
  \bibfield  {author} {\bibinfo {author} {\bibfnamefont {D.}~\bibnamefont {Kedar}}, \bibinfo {author} {\bibfnamefont {Z.}~\bibnamefont {Yao}}, \bibinfo {author} {\bibfnamefont {I.}~\bibnamefont {Ryger}}, \bibinfo {author} {\bibfnamefont {J.~L.}\ \bibnamefont {Hall}}, \ and\ \bibinfo {author} {\bibfnamefont {J.}~\bibnamefont {Ye}},\ }\bibfield  {title} {\enquote {\bibinfo {title} {Synthetic fm triplet for am-free precision laser stabilization and spectroscopy},}\ }\href@noop {} {\bibfield  {journal} {\bibinfo  {journal} {Optica}\ }\textbf {\bibinfo {volume} {11}},\ \bibinfo {pages} {58} (\bibinfo {year} {2024})}\BibitemShut {NoStop}%
\bibitem [{\citenamefont {Robinson}\ \emph {et~al.}(2019)\citenamefont {Robinson}, \citenamefont {Oelker}, \citenamefont {Milner}, \citenamefont {Zhang}, \citenamefont {Legero}, \citenamefont {Matei}, \citenamefont {Riehle}, \citenamefont {Sterr},\ and\ \citenamefont {Ye}}]{robinson2019stabilization}%
  \BibitemOpen
  \bibfield  {author} {\bibinfo {author} {\bibfnamefont {J.~M.}\ \bibnamefont {Robinson}}, \bibinfo {author} {\bibfnamefont {E.}~\bibnamefont {Oelker}}, \bibinfo {author} {\bibfnamefont {W.~R.}\ \bibnamefont {Milner}}, \bibinfo {author} {\bibfnamefont {W.}~\bibnamefont {Zhang}}, \bibinfo {author} {\bibfnamefont {T.}~\bibnamefont {Legero}}, \bibinfo {author} {\bibfnamefont {D.~G.}\ \bibnamefont {Matei}}, \bibinfo {author} {\bibfnamefont {F.}~\bibnamefont {Riehle}}, \bibinfo {author} {\bibfnamefont {U.}~\bibnamefont {Sterr}}, \ and\ \bibinfo {author} {\bibfnamefont {J.}~\bibnamefont {Ye}},\ }\bibfield  {title} {\enquote {\bibinfo {title} {Crystalline optical cavity at 4 k with thermal-noise-limited instability and ultralow drift},}\ }\href@noop {} {\bibfield  {journal} {\bibinfo  {journal} {Optica}\ }\textbf {\bibinfo {volume} {6}},\ \bibinfo {pages} {240} (\bibinfo {year} {2019})}\BibitemShut {NoStop}%
\bibitem [{\citenamefont {Parke}\ and\ \citenamefont {Schioppo}(2025)}]{parke2025stabilization}%
  \BibitemOpen
  \bibfield  {author} {\bibinfo {author} {\bibfnamefont {A.~L.}\ \bibnamefont {Parke}}\ and\ \bibinfo {author} {\bibfnamefont {M.}~\bibnamefont {Schioppo}},\ }\bibfield  {title} {\enquote {\bibinfo {title} {Three hundred microsecond optical cavity storage time and 10- 7 active ram cancellation for 10-19 laser frequency stabilization},}\ }\href@noop {} {\bibfield  {journal} {\bibinfo  {journal} {Optics Letters}\ }\textbf {\bibinfo {volume} {50}},\ \bibinfo {pages} {3405} (\bibinfo {year} {2025})}\BibitemShut {NoStop}%
\bibitem [{\citenamefont {Whittaker}\ \emph {et~al.}(1985)\citenamefont {Whittaker}, \citenamefont {Gehrtz},\ and\ \citenamefont {Bjorklund}}]{whittaker1985RAM}%
  \BibitemOpen
  \bibfield  {author} {\bibinfo {author} {\bibfnamefont {E.~A.}\ \bibnamefont {Whittaker}}, \bibinfo {author} {\bibfnamefont {M.}~\bibnamefont {Gehrtz}}, \ and\ \bibinfo {author} {\bibfnamefont {G.~C.}\ \bibnamefont {Bjorklund}},\ }\bibfield  {title} {\enquote {\bibinfo {title} {Residual amplitude modulation in laser electro-optic phase modulation},}\ }\href@noop {} {\bibfield  {journal} {\bibinfo  {journal} {Journal of the Optical Society of America B}\ }\textbf {\bibinfo {volume} {2}},\ \bibinfo {pages} {1320} (\bibinfo {year} {1985})}\BibitemShut {NoStop}%
\bibitem [{\citenamefont {Wong}\ and\ \citenamefont {Hall}(1985)}]{wong1985servoRAM}%
  \BibitemOpen
  \bibfield  {author} {\bibinfo {author} {\bibfnamefont {N.}~\bibnamefont {Wong}}\ and\ \bibinfo {author} {\bibfnamefont {J.~L.}\ \bibnamefont {Hall}},\ }\bibfield  {title} {\enquote {\bibinfo {title} {Servo control of amplitude modulation in frequency-modulation spectroscopy: demonstration of shot-noise-limited detection},}\ }\href@noop {} {\bibfield  {journal} {\bibinfo  {journal} {Journal of the Optical Society of America B}\ }\textbf {\bibinfo {volume} {2}},\ \bibinfo {pages} {1527} (\bibinfo {year} {1985})}\BibitemShut {NoStop}%
\bibitem [{\citenamefont {Li}\ \emph {et~al.}(2012)\citenamefont {Li}, \citenamefont {Liu}, \citenamefont {Wang},\ and\ \citenamefont {Chen}}]{li2012measurementRAM}%
  \BibitemOpen
  \bibfield  {author} {\bibinfo {author} {\bibfnamefont {L.}~\bibnamefont {Li}}, \bibinfo {author} {\bibfnamefont {F.}~\bibnamefont {Liu}}, \bibinfo {author} {\bibfnamefont {C.}~\bibnamefont {Wang}}, \ and\ \bibinfo {author} {\bibfnamefont {L.}~\bibnamefont {Chen}},\ }\bibfield  {title} {\enquote {\bibinfo {title} {Measurement and control of residual amplitude modulation in optical phase modulation},}\ }\href@noop {} {\bibfield  {journal} {\bibinfo  {journal} {Review of Scientific Instruments}\ }\textbf {\bibinfo {volume} {83}} (\bibinfo {year} {2012})}\BibitemShut {NoStop}%
\bibitem [{\citenamefont {Hall}\ \emph {et~al.}(2015)\citenamefont {Hall}, \citenamefont {Zhang},\ and\ \citenamefont {Ye}}]{hall2015accurateRAM}%
  \BibitemOpen
  \bibfield  {author} {\bibinfo {author} {\bibfnamefont {J.~L.}\ \bibnamefont {Hall}}, \bibinfo {author} {\bibfnamefont {W.}~\bibnamefont {Zhang}}, \ and\ \bibinfo {author} {\bibfnamefont {J.}~\bibnamefont {Ye}},\ }\bibfield  {title} {\enquote {\bibinfo {title} {Accurate removal of ram from fm laser beams},}\ }in\ \href@noop {} {\emph {\bibinfo {booktitle} {2015 Joint Conference of the IEEE International Frequency Control Symposium \& the European Frequency and Time Forum}}}\ (\bibinfo {organization} {IEEE},\ \bibinfo {year} {2015})\ pp.\ \bibinfo {pages} {713--716}\BibitemShut {NoStop}%
\bibitem [{\citenamefont {Zhang}\ \emph {et~al.}(2014)\citenamefont {Zhang}, \citenamefont {Martin}, \citenamefont {Benko}, \citenamefont {Hall}, \citenamefont {Ye}, \citenamefont {Hagemann}, \citenamefont {Legero}, \citenamefont {Sterr}, \citenamefont {Riehle}, \citenamefont {Cole} \emph {et~al.}}]{zhang2014reductionRAM}%
  \BibitemOpen
  \bibfield  {author} {\bibinfo {author} {\bibfnamefont {W.}~\bibnamefont {Zhang}}, \bibinfo {author} {\bibfnamefont {M.}~\bibnamefont {Martin}}, \bibinfo {author} {\bibfnamefont {C.}~\bibnamefont {Benko}}, \bibinfo {author} {\bibfnamefont {J.}~\bibnamefont {Hall}}, \bibinfo {author} {\bibfnamefont {J.}~\bibnamefont {Ye}}, \bibinfo {author} {\bibfnamefont {C.}~\bibnamefont {Hagemann}}, \bibinfo {author} {\bibfnamefont {T.}~\bibnamefont {Legero}}, \bibinfo {author} {\bibfnamefont {U.}~\bibnamefont {Sterr}}, \bibinfo {author} {\bibfnamefont {F.}~\bibnamefont {Riehle}}, \bibinfo {author} {\bibfnamefont {G.}~\bibnamefont {Cole}},  \emph {et~al.},\ }\bibfield  {title} {\enquote {\bibinfo {title} {Reduction of residual amplitude modulation to 1$\times$ 10-6 for frequency modulation and laser stabilization},}\ }\href@noop {} {\bibfield  {journal} {\bibinfo  {journal} {Optics letters}\ }\textbf {\bibinfo {volume} {39}},\ \bibinfo {pages} {1980} (\bibinfo {year} {2014})}\BibitemShut {NoStop}%
\bibitem [{\citenamefont {Krantz}\ \emph {et~al.}(2019)\citenamefont {Krantz}, \citenamefont {Kjaergaard}, \citenamefont {Yan}, \citenamefont {Orlando}, \citenamefont {Gustavsson},\ and\ \citenamefont {Oliver}}]{quantumengineersguide}%
  \BibitemOpen
  \bibfield  {author} {\bibinfo {author} {\bibfnamefont {P.}~\bibnamefont {Krantz}}, \bibinfo {author} {\bibfnamefont {M.}~\bibnamefont {Kjaergaard}}, \bibinfo {author} {\bibfnamefont {F.}~\bibnamefont {Yan}}, \bibinfo {author} {\bibfnamefont {T.~P.}\ \bibnamefont {Orlando}}, \bibinfo {author} {\bibfnamefont {S.}~\bibnamefont {Gustavsson}}, \ and\ \bibinfo {author} {\bibfnamefont {W.~D.}\ \bibnamefont {Oliver}},\ }\bibfield  {title} {\enquote {\bibinfo {title} {{A quantum engineer's guide to superconducting qubits}},}\ }\href {\doibase 10.1063/1.5089550} {\bibfield  {journal} {\bibinfo  {journal} {Appl. Phys. Rev.}\ }\textbf {\bibinfo {volume} {6}},\ \bibinfo {pages} {021318} (\bibinfo {year} {2019})}\BibitemShut {NoStop}%
\bibitem [{\citenamefont {Gao}\ \emph {et~al.}(2021)\citenamefont {Gao}, \citenamefont {Rol}, \citenamefont {Touzard},\ and\ \citenamefont {Wang}}]{practicalGuide_gao2021}%
  \BibitemOpen
  \bibfield  {author} {\bibinfo {author} {\bibfnamefont {Y.~Y.}\ \bibnamefont {Gao}}, \bibinfo {author} {\bibfnamefont {M.~A.}\ \bibnamefont {Rol}}, \bibinfo {author} {\bibfnamefont {S.}~\bibnamefont {Touzard}}, \ and\ \bibinfo {author} {\bibfnamefont {C.}~\bibnamefont {Wang}},\ }\bibfield  {title} {\enquote {\bibinfo {title} {Practical guide for building superconducting quantum devices},}\ }\href@noop {} {\bibfield  {journal} {\bibinfo  {journal} {PRX quantum}\ }\textbf {\bibinfo {volume} {2}},\ \bibinfo {pages} {040202} (\bibinfo {year} {2021})}\BibitemShut {NoStop}%
\bibitem [{\citenamefont {Adisa}\ \emph {et~al.}(2026)\citenamefont {Adisa}, \citenamefont {Lee}, \citenamefont {Cox},\ and\ \citenamefont {Koll{\'a}r}}]{adisa2026pound}%
  \BibitemOpen
  \bibfield  {author} {\bibinfo {author} {\bibfnamefont {I.}~\bibnamefont {Adisa}}, \bibinfo {author} {\bibfnamefont {W.~C.}\ \bibnamefont {Lee}}, \bibinfo {author} {\bibfnamefont {K.~C.}\ \bibnamefont {Cox}}, \ and\ \bibinfo {author} {\bibfnamefont {A.~J.}\ \bibnamefont {Koll{\'a}r}},\ }\bibfield  {title} {\enquote {\bibinfo {title} {Pound-drever-hall method for superconducting-qubit readout},}\ }\href@noop {} {\bibfield  {journal} {\bibinfo  {journal} {Physical Review Letters}\ }\textbf {\bibinfo {volume} {136}},\ \bibinfo {pages} {233601} (\bibinfo {year} {2026})}\BibitemShut {NoStop}%
\bibitem [{\citenamefont {Gao}\ \emph {et~al.}(2008)\citenamefont {Gao}, \citenamefont {Daal}, \citenamefont {Vayonakis}, \citenamefont {Kumar}, \citenamefont {Zmuidzinas}, \citenamefont {Sadoulet}, \citenamefont {Mazin}, \citenamefont {Day},\ and\ \citenamefont {Leduc}}]{gao2008_hanger}%
  \BibitemOpen
  \bibfield  {author} {\bibinfo {author} {\bibfnamefont {J.}~\bibnamefont {Gao}}, \bibinfo {author} {\bibfnamefont {M.}~\bibnamefont {Daal}}, \bibinfo {author} {\bibfnamefont {A.}~\bibnamefont {Vayonakis}}, \bibinfo {author} {\bibfnamefont {S.}~\bibnamefont {Kumar}}, \bibinfo {author} {\bibfnamefont {J.}~\bibnamefont {Zmuidzinas}}, \bibinfo {author} {\bibfnamefont {B.}~\bibnamefont {Sadoulet}}, \bibinfo {author} {\bibfnamefont {B.~A.}\ \bibnamefont {Mazin}}, \bibinfo {author} {\bibfnamefont {P.~K.}\ \bibnamefont {Day}}, \ and\ \bibinfo {author} {\bibfnamefont {H.~G.}\ \bibnamefont {Leduc}},\ }\bibfield  {title} {\enquote {\bibinfo {title} {Experimental evidence for a surface distribution of two-level systems in superconducting lithographed microwave resonators},}\ }\href@noop {} {\bibfield  {journal} {\bibinfo  {journal} {Applied Physics Letters}\ }\textbf {\bibinfo {volume} {92}} (\bibinfo {year} {2008})}\BibitemShut {NoStop}%
\bibitem [{\citenamefont {McRae}\ \emph {et~al.}(2020)\citenamefont {McRae}, \citenamefont {Wang}, \citenamefont {Gao}, \citenamefont {Vissers}, \citenamefont {Brecht}, \citenamefont {Dunsworth}, \citenamefont {Pappas},\ and\ \citenamefont {Mutus}}]{mcrae2020materials}%
  \BibitemOpen
  \bibfield  {author} {\bibinfo {author} {\bibfnamefont {C.~R.~H.}\ \bibnamefont {McRae}}, \bibinfo {author} {\bibfnamefont {H.}~\bibnamefont {Wang}}, \bibinfo {author} {\bibfnamefont {J.}~\bibnamefont {Gao}}, \bibinfo {author} {\bibfnamefont {M.~R.}\ \bibnamefont {Vissers}}, \bibinfo {author} {\bibfnamefont {T.}~\bibnamefont {Brecht}}, \bibinfo {author} {\bibfnamefont {A.}~\bibnamefont {Dunsworth}}, \bibinfo {author} {\bibfnamefont {D.~P.}\ \bibnamefont {Pappas}}, \ and\ \bibinfo {author} {\bibfnamefont {J.}~\bibnamefont {Mutus}},\ }\bibfield  {title} {\enquote {\bibinfo {title} {Materials loss measurements using superconducting microwave resonators},}\ }\href@noop {} {\bibfield  {journal} {\bibinfo  {journal} {Review of Scientific Instruments}\ }\textbf {\bibinfo {volume} {91}} (\bibinfo {year} {2020})}\BibitemShut {NoStop}%
\end{thebibliography}%

\clearpage
\onecolumngrid

\makeatletter
\let\addcontentsline\latex@addcontentsline
\makeatother

\setcounter{figure}{0} 
\setcounter{equation}{0}
\setcounter{section}{0}
\setcounter{tocdepth}{2}

\renewcommand{\figurename}{Figure} 
\renewcommand{\thefigure}{S\arabic{figure}} 
\renewcommand{\thetable}{S\arabic{table}} 
\renewcommand{\theequation}{S\arabic{equation}} 
\renewcommand{\thesection}{S\arabic{section}}

\renewcommand{\theHfigure}{S\arabic{figure}}
\renewcommand{\theHequation}{S\arabic{equation}}
\renewcommand{\theHsection}{S\arabic{section}}

\newpage 
\begin{center}
    \Large Supplemental Material for Robust cavity metrology beyond the limits of Pound-Drever-Hall
\end{center}

\tableofcontents

\vskip 0.6in

\newpage
\section{Observables and Ideal Cavity Response}

\begin{figure}[h!]
    \centering
    \includegraphics[width=0.8\linewidth]{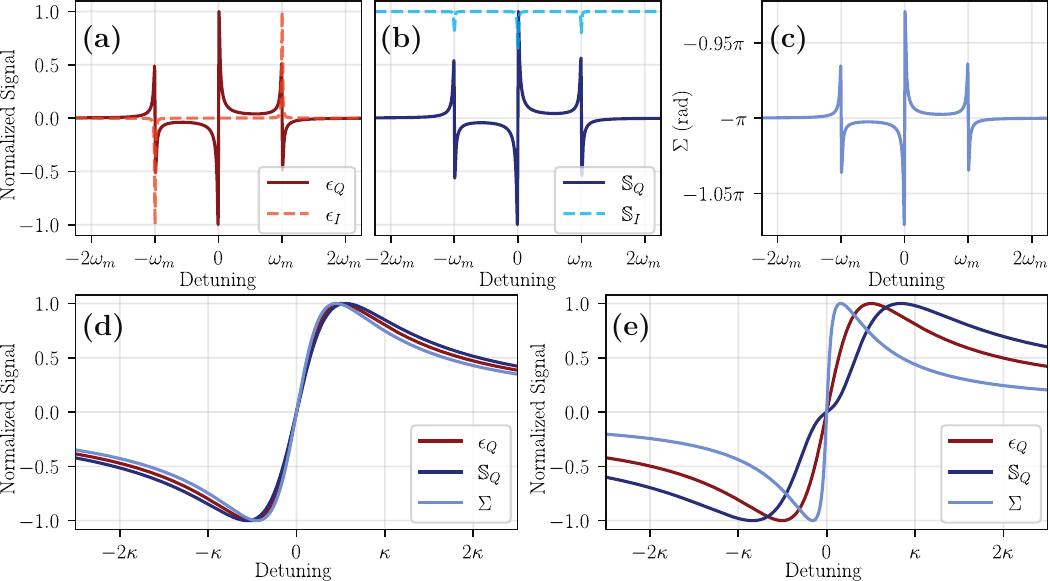}
    \caption{\textit{Ideal three-tone observables.} Ideal lineshapes versus probe detuning for
    (a) Conventional PDH quadratures $\epsilon_{Q}$ and $\epsilon_{I}$,
    (b) Scissors quadratures $\mathbb{S}_{Q}$ and $\mathbb{S}_{I}$,
    (c) Scissors phase $\Sigma$, as the carrier and the two sidebands are swept across the cavity. 
    (d) Normalized comparison of the three observables ($\epsilon_{Q}$, $\mathbb{S}_{Q}$, and $\Sigma$ ) near the cavity resonance in the weak-coupling regime where they all exhibit nearly equivalent dispersive responses. 
    (e) Normalized comparison of the three observables in the strong-coupling regime where $\mathbb{S}_{Q}$ shows a cubic response at the origin while $\epsilon_{Q}$ remains approximately linear and $\Sigma$ retains a sharp phase change associated with resonance. }
    \label{fig:ErrorSignals_highSq}
\end{figure}

The main text introduced three observables constructed from the same three-tone cavity measurement: the conventional PDH signal $\epsilon_Q$, the scissors phase $\Sigma$, and the efficient (arctangent-free) scissors-phase observable $\mathbb{S}_Q$. Conventional PDH provides a sensitive dispersive signal, but its zero-crossing can shift when sideband imbalance is combined with (de)modulation phase errors, as analyzed in Section~\ref{sec:zero_crossing}. The scissors phase rejects these errors directly, although extracting $\Sigma$ from the individual tone phases requires arctangent operations and phase-unwrapping procedures. In contrast, $\mathbb{S}_Q$, which is proportional to $\sin{(\Sigma})$, can be computed directly from the measured IQ components using only addition, subtraction, and multiplication. It requires no arctangent evaluation, division, or phase unwrapping, making it substantially more efficient and robust for real-time implementation, while preserving the error rejection and robust zero-crossing of $\Sigma$.

Here, we derive how $\epsilon_Q$, $\Sigma$, and $\mathbb{S}_Q$ arise from the three-tone probe field and compare their ideal cavity responses. Section~\ref{sec:ideal_beatnote_construction} establishes the description and definition of the measurement observables, Section~\ref{sec:weak_coupling} shows the behavior of the observables in the weak- and strong- coupling regime, respectively. Section~\ref{sec:sq_arctangent_free} then derives the direct IQ construction of $\mathbb{S}_Q$ showing its computational advantage over extracting $\Sigma$. 

\subsection{3-tone probe field and measurement observables} \label{sec:ideal_beatnote_construction}

The conventional PDH and scissors observables extract different combinations of the amplitude and phase information contained in the same three-tone probe field. We write the complex field at the detector as 
\begin{equation}
    \mathbf{E}(t) = \mathbf{A}_{-}e^{i\omega_{-}t} + \mathbf{A}_{0}e^{i\omega_{0}t} + \mathbf{A}_{+}e^{i\omega_{+}t},
\end{equation}
where $\omega_{\pm} = \omega_{0} \pm \omega_{m}$. 
The measured phasors $\mathbf{A}_{j} = E_j e^{i\phi_j}$ and phases $\phi_{j}$ are defined at the detector and therefore include the properties of the generated tone together with the amplitude and phase response acquired from the cavity through the measurement chain. This convention is natural when the three tones are measured individually, but differs from the standard optical PDH convention in which the $\pi$ phase of the lower sideband under ideal phase modulation is often written as an explicit minus sign. Consequently, some expressions below differ in appearance from conventional optical PDH formulas while describing the same physical signal. 

In general, the phase information is carried by the beatnotes generated between the carrier and the sidebands:
\begin{align}
    &\mathbf{B}_{+} = \mathbf{A}_{0}\mathbf{A}_{+}^{*} = E_{0}E_{+}e^{i(\phi_{0} - \phi_{+})}, \\ 
    &\mathbf{B}_{-} = \mathbf{A}_{-}\mathbf{A}_{0}^{*} = E_{0}E_{-}e^{-i(\phi_{0} - \phi_{-})}.
\end{align}
Both the conventional PDH observables and the scissors observables are constructed from $\mathbf{B}_{+}$ and $\mathbf{B}_{-}$. The difference lies in how the two beatnotes are combined. The complex PDH signal $\epsilon$ is the sum of the beatnotes \cite{adisa2026pound}
\begin{equation} \label{eqs:complex_pdh}
    \epsilon = \mathbf{B}_{+} + \mathbf{B}_{-},
\end{equation}
and its two quadratures are
\begin{align}
    \epsilon_{I} = \mathrm{Re}(\epsilon) = E_{+}E_{0}\cos{(\phi_{0} - \phi_{+})} + E_{-}E_{0}\cos{(\phi_{0} - \phi_{-})}, \\
    \epsilon_{Q} = \mathrm{Im}(\epsilon) = E_{+}E_{0}\sin{(\phi_{0} - \phi_{+})} - E_{-}E_{0}\sin{(\phi_{0} - \phi_{-})}.
\end{align}
The PDH signal $\epsilon_{Q}$ is the conventional dispersive PDH error signal. Under ideal phase modulation and the appropriate demodulation phase, $\epsilon_{Q}$ provides a steep central discriminator used for frequency locking~\cite{poundDreverHall1983laser, black2001PDHoverview} and state discrimination~\cite{adisa2026pound} (\fref{fig:ErrorSignals_highSq}a). The complementary quadrature $\epsilon_{I}$ also shown in \fref{fig:ErrorSignals_highSq}a is not normally used as the error signal but becomes important especially when technical phase errors rotate the measurement axis away from the nominal $Q$ axis (see Section~\ref{sec:microwave_IQ_experiment}).  

The scissors phase combines the two carrier-sideband differences without amplitude weighting:
\begin{equation}
    \Sigma = (\phi_0 - \phi_+) + (\phi_0 - \phi_-) = 2\phi_0 - (\phi_- + \phi_+).
\end{equation}
We construct a complex observable whose phase is exactly $\Sigma$ by combining the beatnotes $\mathbf{B}_{+}$ and $\mathbf{B}_{-}$ multiplicatively rather than additively. We define the complex scissors product as
\begin{equation} \label{eqn:supple_S}
    \mathbb{S} = \mathbf{B}_{+}\mathbf{B}_{-}^{*}.
\end{equation}
Substituting the beatnote definition yields
\begin{align}
    \nonumber
    \mathbb{S} &= E_{+}E_{-}E_{0}^{2}e^{i\left[2\phi_{0} - (\phi_{-} + \phi_{+}) \right]}, \\
     &= E_{+}E_{-}E_{0}^{2}e^{i\Sigma}.   
\end{align}
The IQ components of $\mathbb{S}$ are 
\begin{align}
    \mathbb{S}_{I} &= \mathrm{Re}(\mathbb{S}) = E_{0}^{2}E_{+}E_{-}\cos{(\Sigma)}, \\
    \mathbb{S}_{Q} &= \mathrm{Im}(\mathbb{S}) = E_{0}^{2}E_{+}E_{-}\sin{(\Sigma)}.
\end{align}
The quantities $\Sigma$, $\mathbb{S}_{I}$ and $\mathbb{S}_{Q}$ describe the same collective phase coordinate in different forms. The scissors phase $\Sigma$ tracks only the phase $2\phi_{0} - (\phi_{+} + \phi_{-})$, and removes the amplitude scaling in $\mathbb{S}$. The pair ($\mathbb{S}_{I}, \mathbb{S}_{Q}$) instead represents this phase as a point in a cartesian plane whose instantaneous radius is $|\mathbb{S}| = E_{0}^{2}E_{+}E_{-}$. Their responses are shown in \fref{fig:ErrorSignals_highSq}b and the corresponding phase response $\Sigma$ is shown in \fref{fig:ErrorSignals_highSq}c. Although $\mathbb{S}_{Q}$ appears algebraically more complicated than $\Sigma$, Section~\ref{sec:sq_arctangent_free} shows that it can be calculated directly from the measured IQ components in a more efficient way for real-time processing using simple computations. 

The ideal spectral responses of all the observables as the three-tone field is swept across the cavity are shown in \fref{fig:ErrorSignals_highSq}a-c. Each signal contains three principal resonance features corresponding to when the upper sideband, carrier, and the lower sideband pass through the cavity frequency at detunings of $-\omega_{m}$, $0$, and $\omega_{m}$, respectively. The three $Q$-type observables ($\epsilon_{Q}, \Sigma, \mathbb{S}_{Q}$) provide odd, antisymmetric, linear discriminator responses near resonance. The $I$-type observables supply the corresponding orthogonal components. 

\subsection{Line shapes} \label{sec:weak_coupling}
In the weak coupling limit, $\epsilon_{Q}$, $\Sigma$, and $\mathbb{S}_{Q}$ have the same local dependence on the cavity-induced scissors-phase. In this limit, the cavity response is shallow, the detected tone amplitudes vary only weakly across resonance, and the cavity-induced phase shifts are small. Consider an ideal phase modulation where the upper sideband is generated in phase with the carrier ($\phi_{+} = \phi_{0}$), while the lower sideband differs by a $\pi$ phase shift ($\phi_{-} = \phi_{0} + \pi$). Due to the effect of cavity resonance, each tone accumulates small additional phase $\delta\phi$ such that the modified phases become
\[ \phi'_{0} = \phi_{0} + \delta\phi_{0}, \qquad \phi_{+}' = \phi_{+} + \delta\phi_{+}, \qquad \phi_{-}' = \phi_{-}+\delta\phi_{-}.\]
Consequently, the scissors phase becomes
\begin{align}
    \nonumber
    \Sigma &= 2\phi_{0}' - (\phi_{+}' + \phi_{-}')  = 2\phi_{0} - (\phi_{+} + \phi_{-}) + 2\delta\phi_{0} - (\delta\phi_{+} + \delta\phi_{-}) \\
    &= 2\phi_{0} - (\phi_{+} + \phi_{-}) + \delta\Sigma,
\end{align}
where $\delta\Sigma$ is the cavity-induced change in the scissors phase. For ideal phase modulation
\begin{equation}
    \nonumber
    \Sigma = -\pi + \delta\Sigma. 
\end{equation}
Similarly, the PDH signal
\begin{align}
    \nonumber
    \epsilon_{Q} &= E_{+}E_{0}\sin{(\phi_{0}' - \phi_{+}')} - E_{-}E_{0}\sin{(\phi_{0}' - \phi_{-}')} \\
    \nonumber 
    &= E_{+}E_{0}\sin{(\phi_{0} + \delta\phi_{0} - \phi_{+} - \delta \phi_{+})} - E_{-}E_{0}\sin{(\phi_{0} + \delta\phi_{0} - \phi_{-} - \delta\phi_{-})}.
\end{align}
In the case of ideal phase modulation $E_{+} = E_{-} = E_{s}$, and
\begin{equation}
    \epsilon_{Q} = E_{0}E_{s}\sin{(\delta\phi_{0} - \delta\phi_{+})} + E_{0}E_{s}\sin{(\delta\phi_{0} - \delta\phi_{-})},
\end{equation}
which for weak resonances ($|\delta\phi_{0} - \delta\phi_{\pm}| \ll 1 $) this becomes
\begin{align} \label{eqn:supple_pdhq_propto_sigma}
    \nonumber
    \epsilon_{Q} &= E_{0}E_{s}\left[2\delta\phi_{0} - (\delta\phi_{+} + \delta\phi_{-}) \right] \\
    & = E_{0}E_{s}\delta\Sigma.
\end{align}
Therefore, under ideal conditions: ideal phase modulation, perfect demodulation, and weak coupling, conventional PDH measures the scissors-phase ($\epsilon_{Q} \propto \delta\Sigma$,  \eqsref{eqn:supple_pdhq_propto_sigma}). By applying the same phase definitions and limits to $\mathbb{S}_{Q} = E_{0}^{2}E_{+}E_{-}\sin{(\Sigma)}$,
\begin{align}
    \nonumber
    \mathbb{S}_{Q} &= E_{0}^{2}E_{s}^{2}\sin{(-\pi + \delta\Sigma)} \\
    & = -E_{0}^{2}E_{s}^{2}\delta\Sigma.
\end{align}
Thus, apart from the overall scaling constants, the three observables ideally carry the same phase information:
\[ \epsilon_{Q} \propto \delta\Sigma, \quad \Sigma + \pi = \delta\Sigma, \quad \mathbb{S}_{Q} \propto \delta\Sigma. \]
The three signals (in this limit) are shown in \fref{fig:ErrorSignals_highSq}d. The small differences away from the zero-crossing arise from the distinct amplitude weighting and nonlinear phase dependence retained by each observable. These differences become more distinct as the coupling is increased, the resonance becomes deeper, and the carrier amplitude varies strongly. In the strong-coupling limit, the three observables develop distinct central responses: $\Sigma$ becomes narrower and sharper, $\epsilon_Q$ retains an approximately linear discriminator with a characteristic width of $\kappa/2$ (where $\kappa$ is the linewidth of the cavity),  and $\mathbb{S}_Q$ develops a flattened, cubic-like zero-crossing, as shown in \fref{fig:ErrorSignals_highSq}e. This extreme limit corresponds to a resonance for which the magnitude of the scattering coefficient approaches zero. Such a regime is not the primary operating limit for practical qubit-readout and cavity-locking applications, so the reduced local slope of $\mathbb{S}_Q$ is not expected to be a practical limitation for these measurements.

\subsection{Arctangent free construction of $\mathbb{S}_{Q}$} \label{sec:sq_arctangent_free}

To directly compute $\Sigma$, the most obvious implementation is to extract the three phases $\phi_-, \phi_0,$ and $\phi_+$ from the measured IQ data. In practice, such direct evaluation of $\Sigma$ requires multiple evaluations of arctangent which is nonlinear; can become poorly conditioned for noisy IQ points; is subject to phase-wrapping errors; and is slow, leading to latency bottlenecks in FPGA-based feedback loops used for optical cavity locking or in qubit readout.

To bypass these limitations, we construct an experimentally efficient scissors-phase observable $\mathbb{S}_Q=E_{+}E_{-}E^{2}_{0}\sin{\Sigma}$. As we will show here, $\mathbb{S}_Q$ can be computed directly from the measured IQ components using only addition, subtraction, and multiplication. This is closely analogous to the synthetic PDH construction where the quadrature components of the carrier-sideband beatnotes are obtained from dot and cross products of the measured IQ vectors~\cite{adisa2026pound}.

Let the measured IQ vector for tone $j \in \{-, 0, +\}$ be $\mathbf{v}_{j} = (I_{j}, Q_{j}, 0)$. The carrier-sideband beatnotes are
\begin{align} \label{eqs:b_as_dotcross}
    \mathbf{B}_{\pm} &= [\mathbf{v_{\pm}} \cdot \mathbf{v}_{0}]
    \pm i[\mathbf{v_{\pm}} \times \mathbf{v}_{0}]_{\mathrm{z}} \\
    &= (I_0I_\pm + Q_0Q_\pm) \pm i(I_\pm Q_0 - I_0 Q_\pm).
\end{align}
As the complex PDH signal $\epsilon = \epsilon_I + i\epsilon_Q$ is derived from the linear combination of the two carrier-sideband beatnotes (\eqsref{eqs:complex_pdh})
\begin{align}
        \epsilon_{I} &= \mathrm{Re}\left(\mathbf{B}_{+} + \mathbf{B}_{-}\right) = (\mathbf{v}_{+} \cdot \mathbf{v}_{0}) + (\mathbf{v}_{-} \cdot \mathbf{v}_{0}), \\
        \nonumber
        &= \left(I_{0}I_{+} + Q_{0}Q_{+}\right) + \left(I_{0}I_{-} + Q_{0}Q_{-}\right), 
\end{align}
and
\begin{align} \label{eqs:pdh_q_cross-dot}
        \epsilon_{Q} &= \mathrm{Im}\left(\mathbf{B}_{+} + \mathbf{B}_{-}\right) = [\mathbf{v}_{+} \times \mathbf{v}_{0}]_{\mathrm{z}} - [\mathbf{v}_{-} \times \mathbf{v}_{0}]_{\mathrm{z}}, \\
        \nonumber
        &= \left(Q_{0}I_{+} - I_{0}Q_{+}\right) + \left(I_{0}Q_{-} - Q_{0}I_{-}\right).
\end{align}
On the other hand, the scissors observable $\mathbb{S}$ is obtained by multiplying the upper beatnote by the complex conjugate of the lower beatnote: $\mathbb{S}  = \mathbb{S}_{I} + i\mathbb{S}_{Q} = \mathbf{B}_{+}\mathbf{B}_{-}^{*}$ (\eqsref{eqn:supple_S}). Using \eqsref{eqs:b_as_dotcross},
\begin{align}
    \mathbb{S}_{I} &= \mathrm{Re}\left(\mathbf{B}_{+}\mathbf{B}_{-}^{*}\right) = [\mathbf{v}_{+} \cdot \mathbf{v}_{0}][\mathbf{v}_{-} \cdot \mathbf{v}_{0}] - [\mathbf{v}_{+} \times \mathbf{v}_{0}]_{\mathrm{z}}[\mathbf{v}_{-} \times \mathbf{v}_{0}]_{\mathrm{z}}, \\
    \nonumber
    & =  (Q_0Q_+ + I_0I_+)(Q_0Q_- + I_0I_-) - (Q_0I_+ - I_0Q_+)(Q_0I_- - I_0Q_-),
\end{align}
and
\begin{align} \label{eqs:s_q}
    \mathbb{S}_{Q} &= \mathrm{Im}\left(\mathbf{B}_{+}\mathbf{B}_{-}^{*}\right) = [\mathbf{v}_{+} \times \mathbf{v}_{0}]_{\mathrm{z}} [\mathbf{v}_{-} \cdot \mathbf{v}_{0}] + [\mathbf{v}_{-} \times \mathbf{v}_{0}]_{\mathrm{z}} [\mathbf{v}_{+} \cdot \mathbf{v}_{0}], \\
    \nonumber
    & =  (Q_0I_+ - I_0Q_+)(Q_0Q_- + I_0I_+) + (Q_0I_- - I_0Q_-)(Q_0Q_+ + I_0I_+),
\end{align}
are the two quadratures of a beatnote-of-beatnotes, $\mathbb{S}$, whose phase is the scissors phase. This cross and dot product construction allows $\mathbb{S}_Q$ to be evaluated directly from the measured quadratures without first extracting the individual tone phases. This form also simplifies repeated measurements because it can be computed for each sample or single shot and then averaged using an ordinary arithmetic mean. In contrast, extracting $\Sigma$ requires the individual $I$ and $Q$ components to be well resolved before evaluating their phases with an arctangent. In practice, the heterodyne quadratures must therefore be averaged first and $\Sigma$ computed only afterward, requiring the measurement to remain stable over the averaging interval needed to resolve the tone phases. If the phase also crosses an arctangent branch cut, phase unwrapping or circular averaging becomes necessary. However, $\mathbb{S}_Q$ can be computed directly from each noisy measurement and averaged afterward, without first resolving the individual tone phases. This removes the pre-averaging requirement, reducing both computational cost and processing complexity, and makes $\mathbb{S}_Q$ better suited to low-SNR and real-time measurements.

\section{Measurement Setup}
To experimentally validate the ideal observable constructions and systematically investigate the effects of hardware imperfections, we implemented simultaneous multi-tone detection in both optical and superconducting microwave architectures. Both implementations use an additional LO to resolve the three probe tones. In the optical experiment, AOM-generated probe tones and an LO are detected together on a photodiode, which also preserves the conventional PDH beatnote. In the microwave experiment, the IQ-modulated probe tones and LO are combined in a mixer, followed by the simultaneous heterodyne detection and triple digital downconversion architecture developed in~\cite{adisa2026pound}. Despite their different hardware, both measurements produce measurements of the quadratures of each tone, and allow computation of $\epsilon_{Q}$, $\Sigma$, and $\mathbb{S}_{Q}$. 

\subsection{Optical experiment} \label{sec:optical_setup}
\begin{figure} [t]
    \centering
    \includegraphics[width=0.85\linewidth]{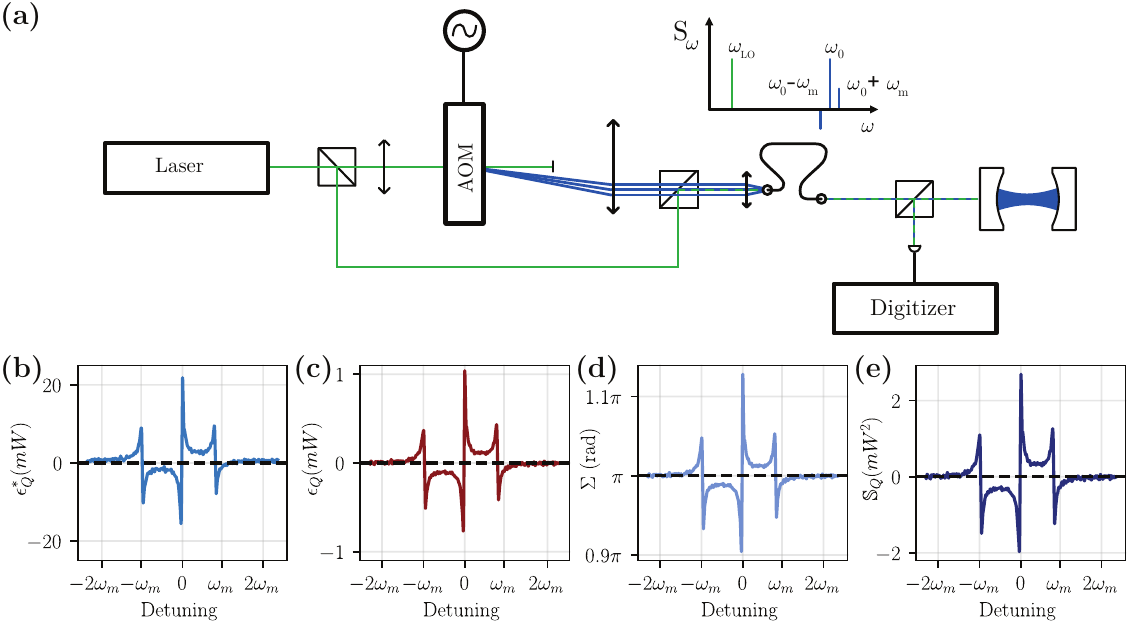}
    \caption{\textit{LO-assisted optical heterodyne detection and reconstruction of the three-tone observables.} 
    (a) Schematic of the optical measurement architecture. The green path denotes the unshifted laser field used as the optical LO. The blue path denotes the AOM-shifted probe field: the AOM is driven by three RF frequencies, producing three first-order diffracted optical tones with independently controlled amplitudes and phases that form the carrier and the two sidebands. The LO and probe tones are combined into a single-mode fiber, shown by the black loop, to place all four tones in the same spatial mode. The far detuned LO is promptly reflected from the cavity, while the three probe tones acquire the cavity response; their interference is detected on a photodiode. 
    (b) Direct conventional PDH error signal $\epsilon_{Q}^{*}$ from the carrier-sideband beatnote generated directly by the photodiode.
    (c) The synthetic PDH signal $\epsilon_{Q}$ reconstructed from the individually acquired heterodyne tones.
    (d) The scissors phase $\Sigma$.
    (e) Arctangent-free scissors-phase observable $\mathbb{S}_{Q}$. 
    }
    \label{fig:supple_opticalSetup}
\end{figure}

A conventional optical PDH measurement does not provide the individual complex amplitudes of the carrier and sidebands required to construct $\Sigma$ and $\mathbb{S}_Q$. Square-law detection followed by demodulation at $\omega_m$ instead combines the two carrier-sideband beatnotes into a single PDH tone. We therefore introduce a far-detuned optical LO that allows the three probe tones to be resolved independently while preserving the conventional PDH beatnote in the same photodetector signal. 

A continuous-wave laser at frequency $\omega_L$ ($\lambda = 556$~nm) is divided into probe and LO paths, as shown in \fref{fig:supple_opticalSetup}a. The probe path passes through an acousto-optic modulator (AOM) driven by an arbitrary waveform generator (Zurich Instruments HDAWG) providing three RF frequencies centered at $\Omega_{\mathrm{AOM}} = 80$~MHz. Each RF component generates a distinct first-order diffracted optical tone, allowing independent control of the amplitude and phase of the carrier and two sidebands. The resulting probe frequencies are
\[ \omega_{j} = \omega_{L} + \Omega_{\mathrm{AOM}} + j\omega_{m}, \quad j\in\{-, 0, +\},\]
for $\omega_{m}=5$~MHz.
Thus, the AOM is used here to synthesize three independently controlled tones, rather than as a conventional single-tone amplitude modulator. The AWG used to drive the AOM ensures precise phase-coherent control of each tone. 


The unshifted laser field bypasses the AOM and serves as an optical LO at $\omega_{LO} = \omega_L$. The LO and the three AOM-generated probe tones are recombined and coupled into a single-mode fiber before reaching the cavity. The fiber places all four tones in the same spatial mode ensuring that they probe the same cavity mode.  
The spatial filtering provided by the fiber comes at the cost of tone-dependent fiber-coupling efficiency, which we compensate for in the AOM drive in order to produce balanced sidebands.
The LO is chosen sufficiently far from the cavity resonance so it is promptly reflected and acts primarily as a detection reference and never comes into resonance with the cavity. This LO-assisted optical architecture is shown schematically in \fref{fig:supple_opticalSetup}a, where the path of the unshifted LO is shown in green, the path of the three AOM-shifted probe tones is denoted in blue, and the loop denotes the single-mode fiber. 

After reflection from the cavity, the detected field can be written as
\begin{equation}
    \mathcal{E}_{det}(t) = a_{LO}e^{i\omega_{LO}t} + \sum_{j=-,0,+}a_{j}e^{i\omega_{j}t},
\end{equation}
where
\[ a_{LO} = E_{LO}e^{i\phi_{LO}}, \quad a_{j}=E_{j}e^{i\phi_j}.\]
The photodiode measures the square-law signal resulting in a photocurrent that contains two distinct sets of beatnotes that are useful for the present measurement. First, beating each tone against the optical LO produces three heterodyne signals at frequencies $\Omega_{j} = \omega_{j} - \omega_{LO} = \Omega_{\mathrm{AOM}} + j\omega_{m}$. These three spectrally-resolved components provide independent access to the amplitude and phase of the lower sideband, carrier, and upper sideband. Second, the carrier beats directly with the two sidebands. The corresponding component at the modulation frequency is the conventional PDH beatnote. We refer to the quadrature extracted from this component as the \emph{direct PDH signal}, denoted by $\epsilon_{Q}^{*}$.  

The output of the photodiode is digitized and digitally demodulated at the four relevant frequencies. Demodulation at $\omega_{m}$ (using $\sin{\omega_{m}t}$) yields the direct PDH signal $\epsilon_{Q}^{*}$, while demolution at $\Omega_{\mathrm{AOM}} + j\omega_{m}$ gives the three complex amplitudes $I_j + i Q_j$. From these complex amplides, $\Sigma$ is obtained from the three detected phases. The synthetic PDH signal $\epsilon_Q$ is reconstructed using the dot- and cross-product formulation (\eqsref{eqs:pdh_q_cross-dot}) developed in Ref.~\cite{adisa2026pound}, while $\mathbb{S}_Q$ is computed using the arctangent-free construction (\eqsref{eqs:s_q} in Sec.~\ref{sec:sq_arctangent_free}). 




When the three probe tones are swept across the cavity, both PDH signals display the same lower-sideband, carrier, and upper-sideband resonance features, together with the same central zero-crossing. Their absolute scales differ because the direct and the reconstructed signals contain different optical-LO powers and electronic gain factors, but their line shapes agree after a demodulation phase calibration with the exception of a slight DC offset in $\epsilon_{Q}^{*}$ due to slight dispersion in the response of the photodiode and RF amplifiers. This general agreement between these signals shows that independent heterodyne measurements of the three tones retains the same differential phase information as the conventional photodetection combined directly at $\omega_{m}$, as shown in \fref{fig:supple_opticalSetup}b and \fref{fig:supple_opticalSetup}c. The constructed $\Sigma$ and $\mathbb{S}_{Q}$, shown in \fref{fig:supple_opticalSetup}d and \fref{fig:supple_opticalSetup}e respectively provide two additional views of the same sweep. The three resonances features remain clearly resolved in all signals, with comparable signal-to-noise ratio to $\epsilon_{Q}^{*}$ and $\epsilon_{Q}$. 

\subsection{Microwave experiment} \label{sec:microwave_setup}

\begin{figure} [t]
    \centering
    \includegraphics[width=0.8\linewidth]{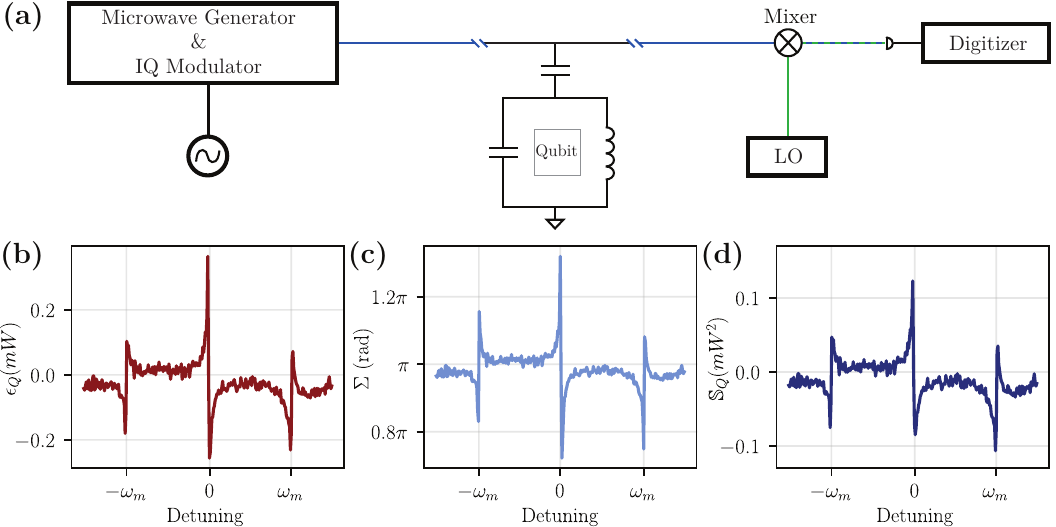}
    \caption{\textit{Microwave triple-downconversion measurement and reconstruction of the three-tone observables.}
    (a) Schematic of the cryogenic microwave measurement architecture. An IQ modulator and a microwave vector generator are used to create the microwave carrier and sideband tones which interact with the qubit-coupled resonator. The transmitted field is mixed with a common LO and digitized, to construct 
    (b) the synthetic PDH signal $\epsilon_{Q}$,
    (c) the scissors phase $\Sigma$, and
    (d) $\mathbb{S}_{Q}$.
    Panels (b) and (c) reproduce the corresponding measurements reported in the Supplemental Material of Ref. \cite{adisa2026pound} and are shown here to enable direct comparison with $\mathbb{S}_{Q}$. }
    \label{fig:supple_microwaveSetup}
\end{figure}

The microwave measurement implements the same observable reconstruction without relying on a cryogenic square-law detector. Instead, the lower sideband, carrier, and upper sideband are measured simultaneously by heterodyne detection using a microwave mixer and separated digitally, using the triple-downconversion or synthetic-PDH architecture we introduced in Ref.~\cite{adisa2026pound} and now extended to construct $\mathbb{S}_{Q}$ from the same measured IQ records. The three probe tones are attenuated and filtered before entering the dilution refrigerator and interacting with the dispersively coupled qubit-resonator system. The outgoing field is amplified by a cryogenic amplification chain, and mixed with a common LO before the three downconverted tones are digitized simultaneously. While the full wiring and component-level implementation are described in Ref.~\cite{adisa2026pound}, the reduced signal path relevant to the reconstruction is shown in~\fref{fig:supple_microwaveSetup}a. Digital demodulation at the three intermediate frequencies corresponding to the three tones gives the three complex amplitudes $I_j + iQ_j$ from which we construct $\epsilon$, $\Sigma$, and $\mathbb{S}$ using the same formulas highlighted above. 

We measure the three observables as a function of carrier detuning while keeping the sideband separation fixed. As the carrier and the sidebands pass the resonator frequency, $\epsilon_Q$, $\Sigma$, and $\mathbb{S}_Q$ display the corresponding carrier and sideband resonance features, as shown in~\ref{fig:supple_microwaveSetup}b - d. The $\epsilon_Q$ and $\Sigma$ measurements previously appeared in Ref.~\cite{adisa2026pound} and are included here to allow direct comparison with $\mathbb{S}_Q$ under the same experimental conditions.

Together, the optical and microwave implementations demonstrate that the three-tone observables can be reconstructed using different physical detection architectures. Both approaches recover the individual complex probe tones and provide access to conventional PDH, the scissors phase, and the arctangent-free scissors observable. This common framework allows the same measurement principles to be applied across continuous optical spectroscopy and discrete microwave-qubit readout.

\section{Effects of Imperfections: robustness of scissors observables}
Having established how the ideal observables are generated and reconstructed in both the optical and microwave measurements, we now account for the physical realities of hardware-induced imperfections in those observables. The main text shows that sideband imbalance and differential-phase errors can shift the conventional PDH zero-crossing, whereas $\Sigma$ and $\mathbb{S}_Q$ reject these technical errors. Here we derive the origin of this distinction and demonstrate it experimentally. The errors acts in two stages: they first change the amplitude- and phase-modulation content of the three-tone probe, and then they affect how the measured tones combine to form $\epsilon_{Q}$, $\Sigma$, and $\mathbb{S}_{Q}$.

These error modes do not occur at the same rate. Common-mode fluctuations $\alpha$, arising from changes in the path length or source-phase drift, are generally the largest and most common. Differential-mode errors $\beta$ are typically smaller and can arise from modulation or demodulation phase errors, frequency-dependent path-length effects, or dispersion across the three tones. Technical scissors-mode deviations $\sigma$ require nonlinear frequency dependence across the modulation bandwidth and are therefore expected to be much rarer. Importantly, $\alpha$, $\beta$, and $\Delta_{\mathrm{amp}}$ are technical changes that cause the measured observable to deviate from the underlying cavity response, whereas $\sigma$ corresponds directly to a change in the scissors-phase itself. 

Section~\ref{sec:modulation_imperfections} describes the imperfect three-tone field in terms of its physical amplitude modulation (AM) and phase-modulation (PM) components. This representation shows directly how sideband imbalance and the collective phase modes alter the intended modulation state. Section~\ref{sec:zero_crossing} then propagates these imperfections into the measurement observables and derives their effects on the discriminator zero-crossings. Sections~\ref{sec:extended_optical_data} and \ref{sec:extended_microwave_data} compare these predictions with controlled optical and microwave measurements respectively.

\subsection{Modulation and sideband generation in the presence of imperfections} \label{sec:modulation_imperfections}
In an ideal PDH measurement, the carrier is accompanied by balanced phase-modulation sidebands. However, hardware imperfections can move this state away from ideal phase modulation by changing the relative amplitudes and phases of the two sidebands. Here we rewrite the imperfections in terms of the collective error modes used in the main text, and give a context to the physical origin of the systematic errors in the measurement observables by tracking the propagation of errors at the field level. The physical origin of the systematic errors is most transparent when the three-tone field is separated into its AM and PM components before constructing any measurement observable. 

We write the three-tone electric field as 
\begin{align} \label{eqn:efield_u+_u-}
    \nonumber
    \mathbf{E}(t) &= E_{-}e^{i(\omega_{0}-\omega_{m})t}e^{i\phi_{-}} + E_{0}e^{i\omega_{0}t}e^{i\phi_{0}} + E_{+}e^{i(\omega_{0}+\omega_{m})t}e^{i\phi_{+}} \\
    \nonumber
    &= e^{i\omega_{0}t}e^{i\phi_{0}} \left[E_{-}e^{i(\phi_{-} - \phi_{0})}e^{-i\omega_{m}t} + E_{0} + E_{+}e^{i(\phi_{+} - \phi_{0})}e^{i\omega_{m}t} \right] \\
    &= e^{i\omega_{0}t}e^{i\phi_{0}} \left[u_{-}e^{-i\omega_{m}t} + E_{0} + u_{+}e^{i\omega_{m}t} \right],
\end{align}
where the sideband phasors are defined relative to the carrier phase as
\begin{equation}
    u_{\pm} = E_{\pm}e^{i(\phi_{\pm} - \phi_{0})}.
\end{equation}
For a general combination of AM and PM, the three-tone field formed by the carrier and first-order sidebands may be written as
\begin{equation} \label{eqn:efield_am_pm}
    \mathbf{E}(t) = e^{i\omega_{0}t}e^{i\phi_{0}} \left[E_{0} + 2A\cos{(\omega_{m}t + \varphi_{A})} + 2iP\sin{(\omega_{m}t + \varphi_{P})} \right],
\end{equation}
where $A$ is the amplitude of the AM-induced sidebands; $P$ is the amplitude of the PM-induced sidebands; and $\varphi_A$ and $\varphi_P$ specify the phases of the modulations. Comparing \eqsref{eqn:efield_am_pm} with the sideband form in \eqsref{eqn:efield_u+_u-} yields
\begin{align}
    u_+ &= Ae^{i\varphi_A} + Pe^{i\varphi_P} \\
    u_- &= Ae^{-i\varphi_A} - Pe^{-i\varphi_P}.
\end{align}
It is then convenient to define the AM phasor as 
\begin{align}
    \nonumber
    \mathbf{A} &= \frac{u_{+}^{*} + u_-}{2} = Ae^{-i\varphi_A}, \\
    \nonumber
    &= A\cos{\varphi_A} - iA\sin{\varphi_A} = A_{c} + iA_{s},
\end{align}
where $A_c$ and $A_s$ are the two AM quadratures. Similarly, we define the PM phasor as 
\begin{align}
    \nonumber
    \mathbf{P} &= \frac{u_{+}^{*} - u_-}{2} = Pe^{-i\varphi_P}, \\
    \nonumber
    &= P\cos{\varphi_A} - iP\sin{\varphi_A} = P_{c} + iP_{s},
\end{align}
where $P_c$ and $P_s$ are the two PM quadratures. We can then rewrite the three-tone electric field in \eqsref{eqn:efield_u+_u-} in terms of the AM and PM quadratures as
\begin{equation}
    \mathbf{E}(t) = e^{i\omega_{0}t}e^{i\phi_{0}} \left[E_{0} + 2A_{c}\cos{\omega_{m}t} + 2A_{s}\sin{\omega_{m}t} - 2iP_{s}\cos{\omega_{m}t}+2iP_{c}\sin{\omega_{m}t} \right].
\end{equation}

We now express these modulation phasors directly in terms of the collective phase coordinates introduced in the main text. The detected phases are decomposed as
\[
(\phi_-, \phi_0, \phi_+) = \mathcal{A}(1,1,1) + \mathcal{B}(-1,0,1) + \frac{\Sigma}{6}\left(-1, 2, -1\right).
\]
The carrier sideband phases are therefore
\begin{equation}
    \phi_+ - \phi_0 = \mathcal{B} - \frac{\Sigma}{2}, \qquad \phi_- - \phi_0 = -\mathcal{B} - \frac{\Sigma}{2}.
\end{equation}
We parametrize the field-amplitude imbalance symmetrically as
\begin{equation}
    E_+ = E_s\left( 1 + \frac{\Delta_{\mathrm{amp}}}{2}\right), \qquad E_- = E_s\left( 1 - \frac{\Delta_{\mathrm{amp}}}{2}\right),
\end{equation}
where $E_s$ is the mean sideband amplitude. The relative sideband phasors become
\begin{align}
    u_+ &= E_s\left( 1 + \frac{\Delta_{\mathrm{amp}}}{2}\right)e^{i(\mathcal{B} - \Sigma/2)}, \\
    u_- &= E_s\left( 1 - \frac{\Delta_{\mathrm{amp}}}{2}\right)e^{-i(\mathcal{B} + \Sigma/2)}.
\end{align}
Using these sideband phasors, we derive the exact AM and PM phasors as
\begin{align}
    \mathbf{A} &= E_s e^{-i\mathcal{B}} \left[ \cos{\left(\frac{\Sigma}{2} \right)} + i \frac{\Delta_{\mathrm{amp}}}{2}\sin{\left(\frac{\Sigma}{2} \right)}\right], \\
    \mathbf{P} &= E_s e^{-i\mathcal{B}} \left[ \frac{\Delta_{\mathrm{amp}}}{2}\cos{\left(\frac{\Sigma}{2} \right)} + i \sin{\left(\frac{\Sigma}{2} \right)}\right].
\end{align}
These expressions separate the roles of the collective phase modes and the sideband amplitudes. The common-mode coordinate $\mathcal{A}$ does not appear because it changes all three absolute phases equally and cancels from relative carrier-sideband phasors. The differential coordinate $\mathcal{B}$ rotates both modulation phasors by the same factor $e^{-i\mathcal{B}}$ i.e. changes the phase of the modulation, but not the kind of modulation. The scissors phase $\Sigma$ and the amplitude imbalance $\Delta_{\mathrm{amp}}$ determine how the field is divided between AM and PM.

In an ideal hardware implementation, the initial state usually lies purely in one of the two bases: AM or PM. For an intended ideal PM state, residual amplitude modulation (RAM) is present whenever the AM phasor is nonzero: $\mathbf{A} \neq 0$. That is, RAM is present whenever at least one of the AM quadratures is nonzero: $A_{c} \neq 0$ or $A_{s} \neq 0$. 
For an ideal PM field with $\phi_0, \phi_+ = 0$ and $\phi_- = -\pi$,
\[
\mathcal{B} = \frac{\pi}{2}, \qquad \Sigma = \pi, \qquad \Delta_{\mathrm{amp}} = 0.
\]
This gives $\mathbf{A} = 0$ and $\mathbf{P} = E_s$, and the field reduces to 
\begin{equation}
    \mathbf{E}_{\mathrm{PM}}(t) = e^{i \omega_{0}t}e^{i\phi_{0}}\left[E_{0} +2iE_{s}\sin{\omega_{m}t} \right].
\end{equation}
Similarly, an ideal AM field with $\phi_-, \phi_0, \phi_+ = 0$ has
\[
\mathcal{B} = 0, \qquad \Sigma = 0, \qquad \Delta_{\mathrm{amp}} = 0,
\]
which gives $\mathbf{A} = E_s$ and $\mathbf{P} = 0$, and 
\begin{equation}
    \mathbf{E}_{\mathrm{AM}}(t) = e^{i \omega_{0}t}e^{i\phi_{0}}\left[E_{0} +2E_{s}\cos{\omega_{m}t} \right].
\end{equation}

We next introduce the phase deviations defined in the main text
\[
(\phi'_-, \phi_0', \phi'_+) = (\phi_-, \phi_0, \phi_+) + \alpha(1,1,1) + \beta(-1,0,1) + \sigma(-1, 2, 1).
\]
The deviations transform the collective coordinates according to
\[
\mathcal{A'} = \mathcal{A} + \alpha, \qquad \mathcal{B'} = \mathcal{B} + \beta, \qquad \Sigma' = \Sigma + 6\sigma.
\]
This decomposition also emphasizes the physical hierarchy of the phase errors. Common-mode fluctuations $\alpha$ are expected to dominate ordinary technical phase noise, while $\beta$ represents smaller differential effects between the sidebands. A technical $\sigma$ requires the carrier phase to change relative to the average sideband phase and is therefore much less common. When $\sigma$ is instead produced by the cavity response, it is not a measurement imperfection but the phase displacement that the scissors observables are designed to detect.

For a field that begins in ideal PM, the imperfect modulation phasors are obtained by setting 
\[
\mathcal{B'} = \frac{\pi}{2} + \beta, \qquad \Sigma' = \pi + 6\sigma.
\]
The exact result is
\begin{align}
    \mathbf{A'}_{\mathrm{PM}} &= E_s e^{-i\beta} \left[ \frac{\Delta_{\mathrm{amp}}}{2}\cos{\left(3\sigma \right)} + i \sin{\left(3\sigma \right)}\right], \\
    \mathbf{P'}_\mathrm{PM} &= E_s e^{-i\beta} \left[ \cos{\left(3\sigma \right)} + i \frac{\Delta_{\mathrm{amp}}}{2}\sin{\left(3\sigma \right)}\right].
\end{align}
These equations show directly how each imperfection modifies the intended PM field. We can isolate the specific physical effects of the errors by examining them individually: 
\begin{itemize}
    \item \textbf{Common-mode error ($\alpha$) only:} For common-mode error only, $\beta = \sigma = \Delta_{\mathrm{amp}} = 0$ resulting in $\mathbf{A}' = 0$ and  $\mathbf{P}' = E_{s}$. Thus, the modulation state remains ideal PM. 

    \item \textbf{Differential phase error ($\beta$) only:} A differential phase error creates no AM component ($\mathbf{A}' = 0$). With $\mathbf{P}' = E_{s}e^{-i\beta}$, the full field becomes $\textbf{E}'(t) = e^{i \omega_{0}t}e^{i\phi_{0}}\left[E_{0} +2iE_{s}\sin{(\omega_{m}t + \beta)} \right]$ where $\beta$ rotates the effective modulation phase. Thus, a differential phase error does not generate physical AM and therefore does not produce RAM by itself. It shifts the RF phase of the PM waveform. 


    \item \textbf{Amplitude imbalance ($\Delta_{\mathrm{amp}}$) only:} In this case, $\beta = \sigma = 0$, creating a real (in-phase) AM component $\mathbf{A}' = E_{s}\Delta_{\mathrm{amp}}/2$, while leaving the PM component unchanged ($\mathbf{P}' = E_s$), resulting in a mixture of PM and AM. 


    \item \textbf{Scissors-phase displacement ($\sigma$) only:} For a scissors-mode displacement, $\beta = \Delta_{\mathrm{amp}} = 0$, converting the real-component PM into an imaginary component of AM:  $\mathbf{A}' = iE_{s}\sin{(3\sigma)}$ and $\mathbf{P}' = E_{s}\cos{(3\sigma)}$. 
    When $\sigma$ is produced by the cavity response, this conversion carries the signal of interest. Any other sources of sigma produce a systematic displacement of the same measured coordinate but such contributions require higher-order frequency-dependent phase shifts and are expected to be much rarer than common- or differential-mode errors. 
\end{itemize}
Qualitatively, $\sigma$ produces a misalignment of the sidebands, while $\Delta_{\mathrm{amp}}$ produces a mismatch in the magnitude of the sidebands. Thus, while they both give rise to RAM, they produce different components of the AM phasor.

The same calculation can be repeated for a field that begins in ideal AM, which transforms as 
\begin{align}
    \mathbf{A'}_{\mathrm{AM}} &= E_s e^{-i\beta} \left[ \cos{\left(3\sigma \right)} + i \frac{\Delta_{\mathrm{amp}}}{2}\sin{\left(3\sigma \right)}\right], \\
    \mathbf{P'}_\mathrm{AM} &= E_s e^{-i\beta} \left[ \frac{\Delta_{\mathrm{amp}}}{2}\cos{\left(3\sigma \right)} + i \sin{\left(3\sigma \right)}\right].
\end{align}
The roles of AM and PM are therefore interchanged between the two initial modulation conventions. Here, a sideband-amplitude imbalance now introduces a real in-phase/cosine-quadrature PM leakage into the AM state, while a differential phase error shifts the phase of the AM waveform. A scissors-phase displacement then generates a sine-quadrature PM.

This calculation shows that the modulation state is itself changed by the same collective errors that later appear in the error signals and measurement observables. The following subsection shows how combinations of these errors produce a false offset in $\epsilon_Q$, while $\Sigma$ and $\mathbb{S}_Q$ remain insensitive to the same leakage mechanism.

\subsection{General error dependence of measurement observables ($\epsilon_{Q}, \Sigma, \mathbb{S}_{Q}$)} \label{sec:zero_crossing}
\begin{figure} [t] 
    \centering
    \includegraphics[width=0.9\linewidth]{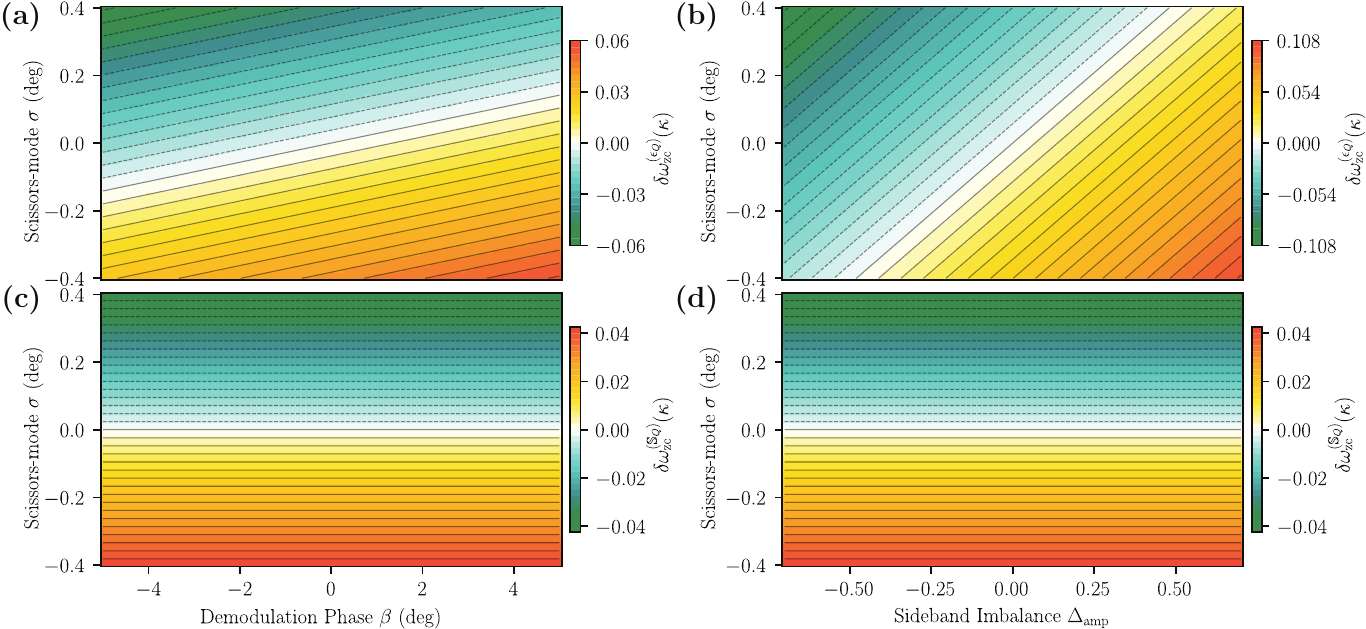}
    \caption{\textit{Zero-crossing displacement of $\epsilon_{Q}$ and $\mathbb{S}_{Q}$ discriminators under combined phase and amplitude errors.}
    (a - b) Calculated resonance-frequency shift $ \delta\omega_{\mathrm{zc}}^{(\epsilon_{Q})}$ for the conventional PDH signal in the presence of scissors-phase displacement $\sigma$ 
    with demodulation phase error $\beta$ varied at fixed sideband amplitude imbalance $\Delta_{\mathrm{amp}}=0.17$ (a), and sideband amplitude imbalance $\Delta_{\mathrm{amp}}$ varied at fixed $\beta=5^\circ$ (b).
    (c-d) Calculated resonance-frequency shift $\delta\omega_{\mathrm{zc}}^{(\mathbb{S}{Q})}$ for $\mathbb{S}{Q}$ under the same conditions: $\beta$ varied at fixed $\Delta_{\mathrm{amp}}=0.17$ (c), and $\Delta_{\mathrm{amp}}$ varied at fixed $\beta=5^\circ$ (d).
    The color scale and contour lines indicate the zero-crossing displacement as a fraction of the cavity linewidth $\kappa$. The PDH zero-crossing shows additional dependence on $\beta$ and $\Delta_{\mathrm{amp}}$, whereas the scissors zero-crossing depends only on $\sigma$, producing contours independent of both $\beta$ and $\Delta_{\mathrm{amp}}$.
    }
    \label{fig:supple_zeroCrossing}
\end{figure}


The impact of systematic errors in a three-tone phase measurement depends on how the signal is used. In discrete state discrimination, such as superconducting-qubit readout, the main concern is rotation of the state-dependent signal away from the chosen measurement axis, which reduces state resolvability. In continuous frequency stabilization, such as optical-cavity locking, the measurement is instead especially sensitive to shifts of the discriminator zero-crossing, since any apparent offset is converted by the feedback loop into a corresponding error in the stabilized frequency. To capture both effects, we first construct the generalized error response of the measurement observables before analyzing their specific failure modes in these two limits. 

Considering the collective phase-errors and sideband-amplitude imbalance as introduced above, the new PDH signal $\epsilon_{Q}'$ in the presence of these errors becomes
\begin{align} \label{eqs:pdh_q'}
    \nonumber
    \epsilon'_{Q} &= E'_{0}E'_{+}\sin{(\phi'_{0} - \phi'_{+})} - E'_{0}E'_{-}\sin{(\phi'_{0} - \phi'_{-})}, \\
    &= E_{0}E_{s}\left(1 + \frac{\Delta_{\mathrm{amp}}}{2}\right)\sin{\left[(\phi_{0}-\phi_{+})-\beta + 3\sigma \right]} - E_{0}E_{s}\left(1 - \frac{\Delta_{\mathrm{amp}}}{2}\right)\sin{\left[(\phi_{0}-\phi_{-})+\beta + 3\sigma \right]}.
\end{align}
To evaluate how these imperfections shift the physical zero crossing, we must determine the compensating cavity detuning required to force the error signal back to zero. Near resonance, the carrier is the tone whose phase changes appreciably with this detuning. The sidebands are assumed to be sufficiently far from resonance that their phases are approximately constant over the small detuning range. Let $\delta\phi_{0}$ be the carrier phase induced by this small cavity detuning. In an ideal error-free system, the zero-crossing occurs exactly at $\delta\phi_{0}=0$. Near resonance, the relative phase differences are $\phi_{0}-\phi_{+} \approx \delta\phi_{0}$ and $\phi_{0}-\phi_{-} = \delta \phi_{0} + \pi$. Assuming the detuning and all phase errors are small ($\delta\phi_{0}, \beta, \sigma \ll 1$), the linearized error response of the PDH signal under the imperfections is
\begin{align} 
    \nonumber
    \epsilon'_{Q}
    &= E_{0}E_{s}\left(1 + \frac{\Delta_{\mathrm{amp}}}{2}\right)\left(\delta\phi_{0}-\beta + 3\sigma \right) + E_{0}E_{s}\left(1 - \frac{\Delta_{\mathrm{amp}}}{2}\right)\left(\delta\phi_{0}+\beta + 3\sigma \right) \\
    \label{eqs:pdh_linear}
    &= E_{0}E_{s}\left[2 \delta\phi_{0} + \left(6\sigma -\beta \Delta_{\mathrm{amp}}\right)\right],
\end{align}
which has the form  $\epsilon'_{Q} = m \delta \phi_{0} + \delta\epsilon_{Q}$. Here $m = 2E_{0}E_{s}$ is the effective discriminator slope and $\delta\epsilon_{Q}=E_{0}E_{s}\left(6\sigma - \beta \Delta_{\mathrm{amp}} \right)$ is the vertical offset of the PDH signal at the ideal zero-crossing due to the errors. We then derive the true zero-crossing displacement (a horizontal frequency offset) by solving $\epsilon'_{Q} = 0$ in \eqsref{eqs:pdh_linear}.

The needed solution to \eqsref{eqs:pdh_linear}, which is the phase shift required to zero the PDH signal is
\begin{equation}
    \delta \phi_{0, \mathrm{zc}}^{(\epsilon_{Q})} = -3\sigma + \frac{\beta\Delta_{\mathrm{amp}}}{2 },
\end{equation}
and the corresponding frequency displacement $\delta\omega_{\mathrm{zc}}^{(\epsilon_{Q})}$ is obtained from the local carrier phase slope $d\phi_0 / d\omega$ since $\delta \phi_{0}$ = ($d\phi_0 / d\omega$) $\delta \omega$ for small detuning. Therefore
\begin{equation} \label{eq:pdh_zero-crossing}
    \delta\omega_{\mathrm{zc}}^{(\epsilon_{Q})} = -\left( \left.\frac{d\phi_{0}}{d\omega} \right|_{\omega_0}\right)^{-1}\left[ 3\sigma - \frac{\beta\Delta_{\mathrm{amp}}}{2 } \right] 
\end{equation}
is the actual zero-crossing of the imbalanced PDH discriminator. 
The two terms in \eqsref{eq:pdh_zero-crossing} have different physical meanings. The $\beta\Delta_{\mathrm{amp}}$ contribution is a technical measurement error: changes in the modulation or detection hardware shift the PDH zero-crossing even though the underlying cavity response has not changed. By contrast, the $-3\sigma$ term represents a displacement of the scissors phase itself. Technical contributions to $\sigma$ are expected to be rare, while a cavity-induced $\sigma$ is precisely the signal that the measurement is intended to track.
The final form of $d\phi_0 / d\omega$ depends on the scattering coefficient $\mathbf{S}(\omega)$ of the cavity/resonator. In general, 
\begin{equation}
    \left.\frac{d\phi_{0}}{d\omega} \right|_{\omega_0} = \left. \frac{d}{d\omega} \mathrm{arg} \left[\mathbf{S(\omega)}\right] \right|_{\omega_{0}} = \frac{\mathcal{D}}{\kappa},
\end{equation}
where $\kappa$ is the linewidth of the cavity, and $\mathcal{D}$ is a dimensionless phase-slope factor that depends on the exact cavity and coupling port geometry. For an hanger resonator~\cite{gao2008_hanger, mcrae2020materials} commonly used for superconducting-qubit readout and used to generate the theory plots in \fref{fig:supple_zeroCrossing},
\begin{equation}
    \mathbf{S}(\omega) = 1 - \frac{1}{\frac{Q_{\mathrm{ext}}}{Q_{\mathrm{int}}} + 1 + 2iQ_{\mathrm{ext}}\frac{\omega - \omega_{0}}{\omega_{0}}},
\end{equation}
for internal and external quality factors $Q_{\mathrm{int}}$ and $Q_{\mathrm{ext}}$, consequently
\begin{equation}
    \left.\frac{d\phi_{0}}{d\omega} \right|_{\omega_0} = \frac{2Q_{\mathrm{int}}Q}{\omega_{0}Q_{\mathrm{ext}}} = \frac{2\kappa_{\mathrm{ext}}}{\kappa \times \kappa_{\mathrm{int}}}.
\end{equation}
Here, $\kappa_{ext}$ is the coupling rate to the feedline and $\kappa_{int}$ is the internal loss rate. Thus
\begin{equation}
    \delta\omega_{\mathrm{zc, hanger}}^{(\epsilon_{Q})} = -\frac{\kappa \times \kappa_{\mathrm{int}}}{2\kappa_{\mathrm{ext}}}\left[ 3\sigma - \frac{\beta\Delta_{\mathrm{amp}}}{2 } \right].
\end{equation}

From $\delta\omega_{\mathrm{zc}}^{(\epsilon_{Q})}$ in \eqsref{eq:pdh_zero-crossing}, the technical zero-crossing error arises from the coupled dependence on $\beta$ and $\Delta_{\mathrm{amp}}$. A differential phase error $\beta$ alone rotates the two carrier-sideband beatnotes but leaves the Q component of their balanced sum centered at the correct resonance. Sideband imbalance breaks this cancellation by weighting the two beatnotes unequally, allowing the phase rotation to produce a false frequency shift. Thus, varying $\beta$ at fixed $\Delta_{\mathrm{amp}}$, or varying $\Delta_{\mathrm{amp}}$ at fixed $\beta$, changes the carrier-phase displacement $\delta\phi_0$ required to recover the PDH zero-crossing, producing the tilted contours in \fref{fig:supple_zeroCrossing}a and \fref{fig:supple_zeroCrossing}b, respectively. In contrast, the $-3\sigma$ contribution shifts the zero-crossing independently of these technical errors because $\sigma$ represents a displacement of the scissors phase itself which is the quantity the measurement is intended to detect.

\noindent \textit{Scissors-Phase Observables}.--- Based on the pure phase coordinates, the scissors phase is defined as 
\begin{align}
    \nonumber
    \Sigma = 2\phi_0 - (\phi_- + \phi_+).
\end{align}
Under the same collective phase errors,
\begin{align}
    \nonumber
    \Sigma' &= 2\phi'_0 - (\phi'_- + \phi'_+) \\
            &= \Sigma + 6\sigma. 
\end{align}
The cancellation of the common-mode ($\alpha$) and the differential-mode ($\beta$) errors is the essential distinction between the scissors phase and the conventional PDH signal. $\Sigma$ is not amplitude-weighted and is thus insensitive to sideband-amplitude imbalance. Near resonance, taking the same limits highlighted above in the case of PDH,  
\begin{equation} \label{eqn:supple_sigma_prime}
    \Sigma' = 2\delta\phi_{0} + \pi + 6\sigma.
\end{equation}
Because the ideal baseline of $\Sigma$ is $\pi$, the phase shift $\delta\phi_{0, \mathrm{zc}}^{(\Sigma)}$ required to make $\Sigma'$ return to its ideal baseline is obtained by solving $\pi = 2\delta\phi_{0, \mathrm{zc}}^{(\Sigma)} + \pi + 6\sigma$ yielding $\delta\phi_{0, \mathrm{zc}}^{(\Sigma)} = -3\sigma$. The corresponding frequency displacement is simply 
\begin{align}
    \delta\omega_{\mathrm{zc}}^{(\Sigma)} = -3\sigma\left( \left.\frac{d\phi_{0}}{d\omega} \right|_{\omega_0}\right)^{-1}.
\end{align}

Similarly for $\mathbb{S}_{Q} = E_{-}E_{+}E_{0}^{2}\sin{\Sigma}$, under amplitude imbalance and collective phase errors,
\begin{align}
    \nonumber
    \mathbb{S}'_{Q} &= E_{s}^{2}\left(1+ \frac{\Delta_{\mathrm{amp}}}{2}\right)\left(1- \frac{\Delta_{\mathrm{amp}}}{2}\right)E_{0}^{2}\sin{(2\delta\phi_{0} + \pi + 6\sigma)}, \\
    &\approx -E_{s}^{2}\left(1+ \frac{\Delta_{\mathrm{amp}}}{2}\right)\left(1- \frac{\Delta_{\mathrm{amp}}}{2} \right)E_{0}^{2}(2\delta\phi_{0} + 6\sigma),
\end{align}
from which we derive $\delta\phi_{0, \mathrm{zc}}^{(\mathbb{S_{Q})}} = -3\sigma$ and
\begin{align}
    \delta\omega_{\mathrm{zc}}^{(\mathbb{S_{Q}})} = -3\sigma\left( \left.\frac{d\phi_{0}}{d\omega} \right|_{\omega_0}\right)^{-1}.
\end{align}

While $\mathbb{S}_Q$ is amplitude-weighted, the amplitudes only change the overall scale and do not move its zero-crossing. Unlike conventional PDH, there is no $\beta\Delta_{\mathrm{amp}}$ offset term. The zero-crossings of $\Sigma$ and $\mathbb{S}_{Q}$ are dictated purely by $\sigma$ i.e. changes in the scissors phase itself, completely immune to modulation phase miscalibrations and amplitude imbalance. This produces perfectly horizontal contours in \fref{fig:supple_zeroCrossing}c and \fref{fig:supple_zeroCrossing}d, in direct contrast to the tilted and curved PDH contours above. For the scissors observables, a plot containing only $\beta$ and $\Delta_{\mathrm{amp}}$ at fixed $\sigma$ would be uniform and is therefore not included.

Furthermore, while the analytical formulas demonstrate that pure differential phase errors $\beta$ acting on a perfectly balanced field do not shift the PDH zero-crossing, it is important to emphasize that they do change the discriminator slope. This reduces the sensitivity to small shifts from resonance and is therefore particularly important for applications such as superconducting-qubit readout, where the qubit state is inferred from a small state-dependent shift of the resonator frequency. Also for continuous frequency locking, the suppression of the discriminator slope results in reducing feedback gain and tracking stability. Thus a differential phase error can still degrade readout performance; and an unchanged zero-crossing does not guarantee a useful PDH discriminator. A differential phase offset physically rotates the measurement axis in the complex plane, continuously mixing the dispersive PDH signal $\epsilon_{Q}$
into $\epsilon_I$~\cite{adisa2026pound}
\[ \epsilon_{Q}' = \epsilon_{Q}\cos{\beta} - \epsilon_I \sin\beta.\]
At $\beta = \pi/2$, the $Q$ quadrature vanishes entirely. A feedback controller using that fixed quadrature then has no restoring signal, despite the persistence of the zero-crossing. While a static rotation can be compensated by changing the demodulation phase, an unknown or fluctuating $\beta$ produces a drifting loop gain and can render the locking ineffective.  For discrete superconducting microwave qubit readout, this quadrature rotation is equally destructive. When the ideal measurement lies in $\epsilon_{Q}$, a static differential phase error rotates the measurement signal away from the readout axis, while shot-to-shot fluctuations in $\beta$ broaden or smear the IQ distributions and reduce state resolvability (see Section \ref{sec:microwave_IQ_experiment}). The scissors observables are immune to $\beta$ fluctuations.

\subsection{Extended Optical Data} \label{sec:extended_optical_data}

\begin{figure} [t]
    \centering
    \includegraphics[width=0.9\linewidth]{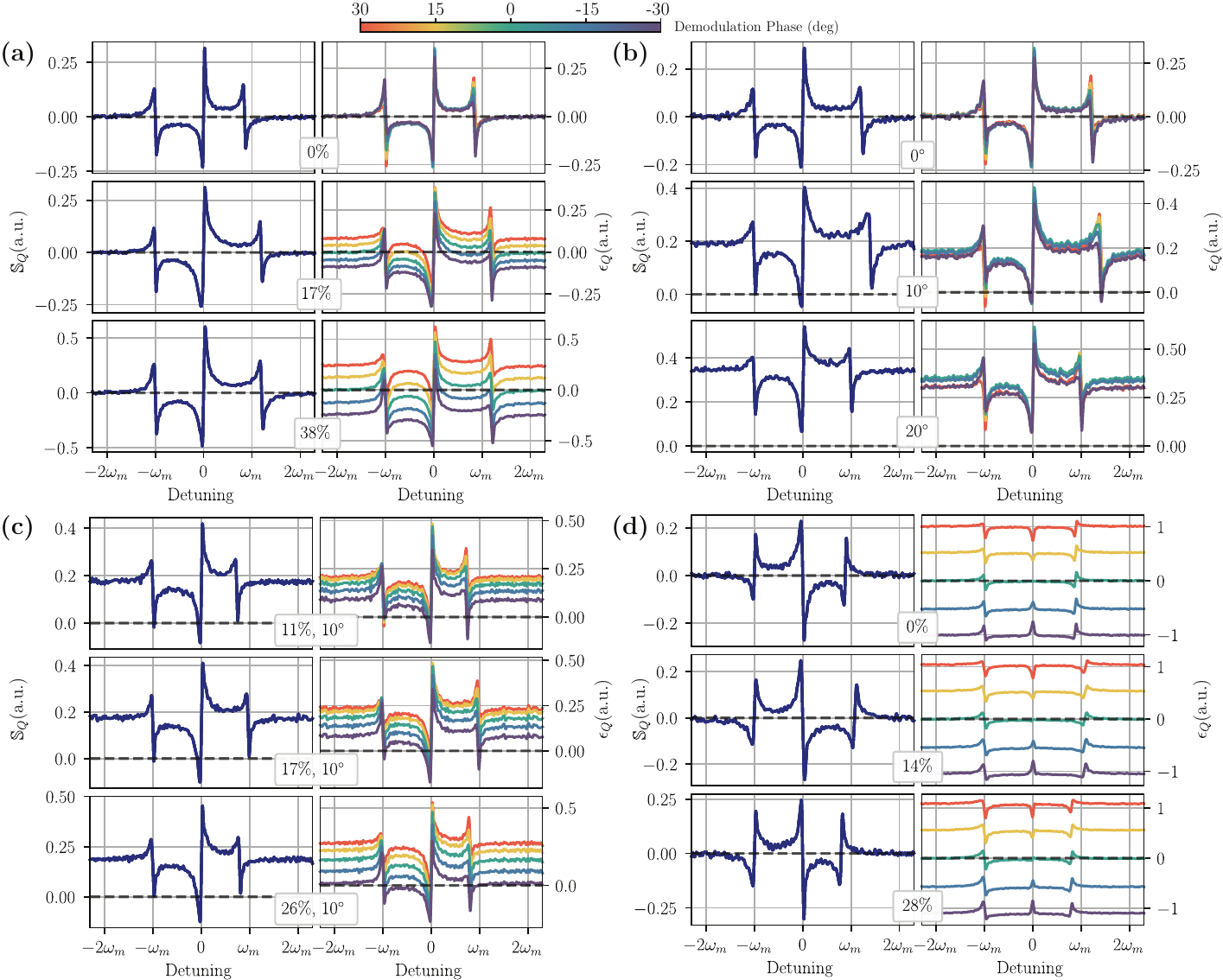}
    \caption{\textit{Optical response of PDH signal $\epsilon_Q$ and scissors-phase observable $\mathbb{S}_{Q} $ to controlled modulation imperfections.}
    Measured $\mathbb{S}_{Q}$ signal (left column of each panel) and PDH signal (right column) as the input signal is swept across the cavity. Trace color denotes the applied demodulation phase, varied from $-30^\circ$ to $30^\circ$.
    (a) Phase-modulated input with sideband power-imbalance increased from $0\%$ to $38\%$. The imbalance changes the amplitude of $\mathbb{S}_{Q}$ without shifting its central zero-crossing, whereas $\epsilon_{Q}$ suffers severe $\beta$-dependent zero-crossing shifts with increasing sideband imbalance.
    (b) Approximately balanced sidebands with an increasing scissors-phase displacement of $0^\circ-20^\circ$. Both observables track the imposed phase displacement but the zero-crossing of $\epsilon$ acquires a weak $\beta$ dependence since $\Delta_{\mathrm{amp}}$ is not absolutely zero. 
    (c) Fixed $10^\circ$ scissors-phase displacement with amplitude imbalance increased from $11\%$ to $26\%$. The zero-crossing of $\mathbb{S}_Q$ remains fixed by the imposed scissors-phase, whereas the zero-crossing of $\epsilon_Q$ acquires $\beta$-dependent shifts. 
    (d) Amplitude-modulated input with sideband power-imbalance increased from $0\%$ to $28\%$. The conventional PDH signal loses its discriminator signal and becomes strongly dependent on the selected demodulation phase, while $\mathbb{S}_{Q}$ retains a well-defined robust zero-crossing discriminator. Small variations in the apparent sideband-resonance positions across the scans arise from mechanical variations during the unlocked cavity sweeps, and do not reflect any observable instability. All traces in each panel are normalized by the average off-resonant value of $E_{+}E_{-}E_{0}^{2}$ (for $\mathbb{S}_{Q}$) or $E_{0}(E_{-}+E_{+})/2$ (for $\epsilon_Q$).
    }
    \label{fig:supple_optical_imbalances}
\end{figure}

Having established the theoretical vulnerability of the PDH readout to amplitude and phase errors, we experimentally validate the stability of $\mathbb{S}_{Q}$ using the optical LO assisted setup described in Section~\ref{sec:optical_setup}. We demonstrate the predicted error sensitivities by independently controlling the sideband imbalance $\Delta_{\mathrm{amp}}$ and scissors-phase displacement $\sigma$ while varying the differential phase $\beta$ for each configuration. In practice, $\beta$ represents an error in the modulation or demodulation phase. For each fixed combination of $\Delta_{\mathrm{amp}}$ and $\sigma$, we sweep the unlocked cavity resonance across the probe tones, record the heterodyne signals, and reconstructed $\mathbb{S}_Q$ and $\epsilon_Q$ as functions of carrier detuning.

Each panel of (\fref{fig:supple_optical_imbalances}) compares $\mathbb{S}_Q$
in the left column with $\epsilon_Q$ in the right column. Different rows correspond to fixed values of $\Delta_{\mathrm{amp}}$ and $\sigma$ while the colored traces show $\beta$ varied from $-30^\circ$ to $30^\circ$. The percentages shown in the figure denote the applied sideband-power imbalance; $\Delta_{\mathrm{amp}}$ is the corresponding field-amplitude imbalance. This arrangement reveals both changes in the signal line shape and shifts in the zero-crossings $\delta\omega_{\mathrm{zc}}^{(\mathbb{S}_Q)}$ and $\delta\omega_{\mathrm{zc}}^{(\epsilon_Q)}$. We observe the effect of the experimental imperfections in four operational regimes in \fref{fig:supple_optical_imbalances}a-d determined by which of $\Delta_{\mathrm{amp}}$ and $\sigma$ that is being controlled with varying $\beta$:

\begin{itemize}
    \item \textbf{Sideband imbalance} $\Delta_{\mathrm{amp}}$:~(\fref{fig:supple_optical_imbalances}a): While keeping $\sigma$ near zero, we first investigate the combined effect of sideband imbalance and demodulation phase error $\beta$. We vary the sideband power imbalance from $0\%$ (perfectly balanced sidebands) up to $38\%$ while simultaneously changing the demodulation phase from $-30^\circ$ to $30^\circ$. When the sidebands are balanced, the opposite phase rotations of the two PDH beatnotes cancel at the zero crossing, although their projection onto the measured quadrature changes. As the imbalance is increased, $\epsilon_{Q}$ develops a phase-dependent offset and moving zero-crossing. However, $\mathbb{S}_{Q}$ behaves differently: its amplitude varies with the sideband powers but the zero-crossing remains completely robust against demodulation phase variations regardless of the severity of the power imbalance. 

    \item \textbf{Scissors-phase displacement} $\sigma$~(\fref{fig:supple_optical_imbalances}b): Next, we isolate the effect of $\sigma$ by introducing a phase offset to both sidebands relative to the carrier, effectively breaking the $\Sigma = \pm\pi$ symmetry condition, but maintaining approximately balanced sidebands. The offset of $\mathbb{S}_{Q}$ moves systematically as the imposed phase displacement is increased, while remaining insensitive to the simultaneous demodulation-phase. This behavior demonstrates that $\mathbb{S}_Q$ retains sensitivity to the phase displacement that the measurement is designed to detect while rejecting $\beta$. The PDH signal also tracks the imposed phase displacement but acquires a weak $beta$ dependence since $\Delta_{\mathrm{amp}}$ is not absolutely zero.

    \item \textbf{Combined sideband imbalance and scissors-phase displacement}~(\fref{fig:supple_optical_imbalances}c): Two effects are now combined by fixing $\sigma$ at $10^\circ$ and varying the sideband power imbalance from $11\%$ to $26\%$. The measured $\mathbb{S}_{Q}$ retains the zero-crossing established by the fixed phase displacement across the full sideband imbalance range, even as the amplitudes change. In contrast, $\epsilon_Q$ acquires additional $\beta$-dependent shifts because the demodulation error couples with the sideband imbalance creating a zero-crossing offset. This result illustrates an important practical distinction: $\mathbb{S}_Q$ continues to report the imposed collective phase displacement, whereas the PDH zero-crossing no longer provides an unambiguous measure of the displacement without independent calibration of the sideband imbalance and demodulation phase. 

    \item \textbf{Amplitude modulation}~(\fref{fig:supple_optical_imbalances}d): Finally, we replace the usual phase-modulated input with an amplitude-modulated three-tone field where the sidebands are generated in-phase with the carrier. Ideal amplitude modulation has $\Sigma = 0$, rather than the $\Sigma = \pm\pi$ of ideal phase modulation. This condition removes the antisymmetric sideband relationship on which the standard PDH discriminator relies. Consequently, the $\epsilon_{Q}$ discriminator signal is completely suppressed, failing to produce a zero-crossing at resonance. However, the $\pi$ difference in $\Sigma$ only reverses the sign of $\mathbb{S}_Q$. Thus it inherently accommodates the modulation condition and continues to measure the carrier phase relative to the sideband average, producing a well-defined steep, robust zero-crossing discriminator. 
\end{itemize}

The cavity is left unlocked and swept across the fixed probe tones, and the detuning is inferred from the elapsed time relative to the carrier resonance using the known sideband spacing. Thus, the horizontal axis is determined from the timing of the cavity sweep rather than from an independently calibrated frequency sweep. Mechanical vibrations of the optical cavity introduce shot-to-shot variations in this timing, which appear as small shifts in the apparent positions of the sideband resonances between scans. These variations reflect the cavity motion rather than any fundamental instability of the observables and do not affect the simultaneous comparison of $\epsilon_{Q}$ and $\mathbb{S}_{Q}$ within a given scan or the dependence of their measured zero-crossings on the deliberately applied errors.

\subsection{Extended microwave data} \label{sec:extended_microwave_data}

\begin{figure} [t]
    \centering
    \includegraphics[width=0.9\linewidth]{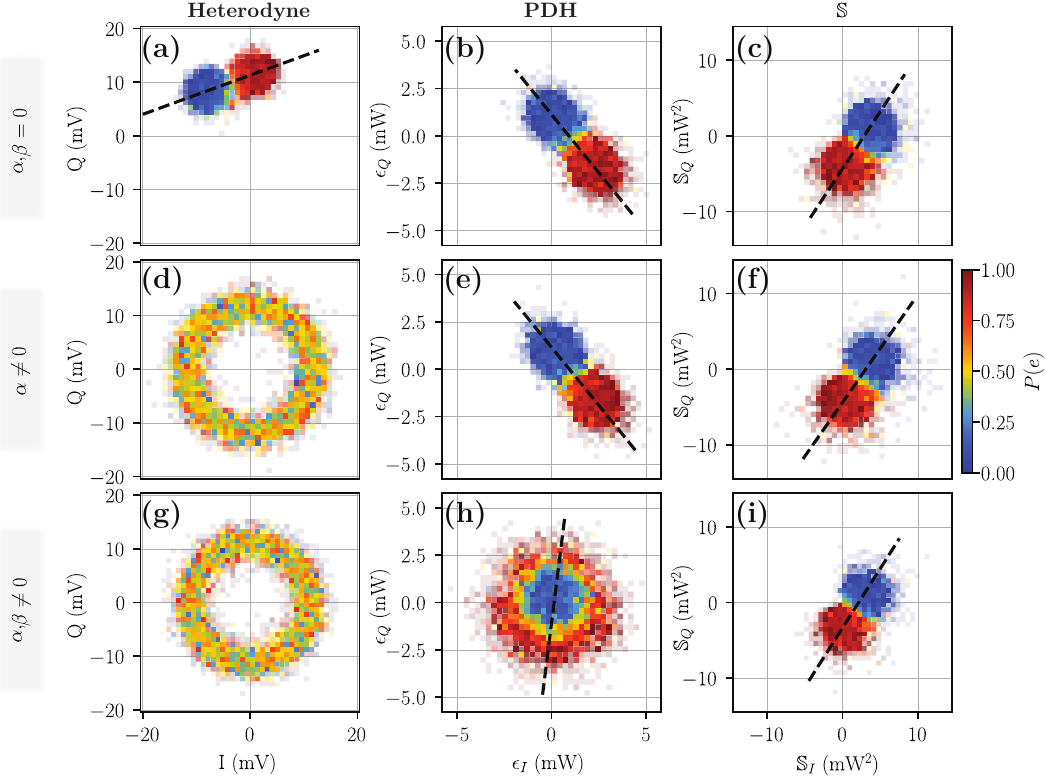}
    \caption{\textit{Microwave qubit-state readout under controlled phase errors}
    (a-c) Single-shot IQ distributions for conventional carrier-heterodyne readout, synthetic PDH readout, and scissors readout under ideal, phase-locked conditions. The optimal projection axis for state discrimination is represented by the dashed black lines and the color scale indicates the probability $P(e)$ of events corresponding to preparation in $\ket{e}$. The opacity is scaled by the total bin occupancy, with bins containing fewer than 10 counts displayed with proportionally reduced opacity.
    (d-f) Readout distributions in the presence of common-mode phase drifts induced by an unlocked carrier generator ($\alpha \neq 0$). Heterodyne no longer yeilds any appreciable readout fidelity, while PDH and $\mathbb{S}$ are unaffected.
    (g-i) Readout distributions under combined common-mode drifts and differential-mode timing errors injected via the arbitrary waveform generator and digitizer ($\alpha, \beta \neq 0$). PDH is partially degraded by random rotation between $\epsilon_I$ and $\epsilon_Q$ while $\mathbb{S}$ remains unaffected. All datasets include $5000$ shots with the qubit prepared in $\ket{g}$ and $5000$ shots with the qubit prepared in $\ket{e}$.}
    \label{fig:supple_microwave_IQs}
\end{figure}

Here we demonstrate the stability of $\mathbb{S}_{Q}$ and show that it preserves qubit-state information when technical phase fluctuations are larger than the phase shift produced by the qubit itself. We compare conventional carrier heterodyne readout, synthetic PDH readout, and scissors readout using the same single-shot records while progressively introducing the collective phase errors developed above. By deliberately introducing the hardware imperfections, we can track the sequential failure of the different measurement observables and demonstrate the superior robustness of $\mathbb{S}_{Q}$.

\subsubsection{Robustness in the IQ plane} \label{sec:microwave_IQ_experiment}
To visualize the state discrimination capability under various error conditions, we construct two-dimensional state-fraction maps in the complex IQ plane, as shown in \fref{fig:supple_microwave_IQs}, using the same plotting convention used in \cite{adisa2026pound}. To display both state separation and overlap within a single IQ map, each two-dimensional pixel is colored by the measured excited state fraction 
\[ P(e) = \frac{N_e}{N_g + N_e},\]
where $N_g$ and $N_e$ are the numbers of shots obtained after preparing the qubit in $\ket{g}$ and $\ket{e}$ respectively. Blue regions contain predominantly ground-state preparations, red regions contain predominantly excited state preparations, and the overlap regions are indicated by the yellow bins. To prevent shot noise in sparsely populated regions from being visually overemphasized, the transparency of each bin is scaled by its total occupancy according to $\mathrm{min}\left[(N_g + N_e)/10 , 1 \right]$~\cite{adisa2026pound}. Thus, bins with fewer than 10 total counts are displayed with reduced opacity, while bins with $N_g+N_e\ge 10$ are shown at full opacity. The carrier IQ coordinates $(I_0, Q_0)$, the complex PDH coordinates $(\epsilon_I , \epsilon_Q)$, and the complex scissors coordinates $(\mathbb{S}_{I}, \mathbb{S}_{Q})$ are shown in the three columns of \fref{fig:supple_microwave_IQs} respectively.

When the microwave generators and digitizer are locked, all three observables yield well-resolved, distinct clusters corresponding to the ground and excited states, and provide a fixed axis along which the two qubit states can be distinguished (\fref{fig:supple_microwave_IQs}a-c). By unlocking the carrier generator from the external reference clock, we introduce severe common-mode phase drifts ($\alpha$) into the system. This destroys the phase reference of the conventional heterodyne signal, collapsing the state clusters into overlapping, indistinguishable rings, as shown in \fref{fig:supple_microwave_IQs}d. Conversely, both PDH (\fref{fig:supple_microwave_IQs}e) and scissors readout (\fref{fig:supple_microwave_IQs}f) reject this absolute phase drift because they depend only on phase differences among the three copropagating tones, cleanly preserving states discrimination. The distinction between PDH ($\epsilon$) and scissors readout ($\mathbb{S}$) becomes apparent when deliberate timing offsets are applied to the arbitrary waveform generator and digitizer, in addition to the unlocked generator, thus injecting both common-mode drifts and differential-mode phase errors. The qubit states remain collapsed for the heterodyne readout (\fref{fig:supple_microwave_IQs}g) while for PDH (\fref{fig:supple_microwave_IQs}h), the $\beta$ error removes the fixed PDH discrimination axis. The PDH outcomes rotate around the origin, and the qubit-state information no longer remains aligned with a single linear quadrature. While some radial information remains in the full two-dimensional distribution, the overall separability is reduced. By contrast, the scissors signal $(\mathbb{S}_{I}, \mathbb{S}_{Q})$ retains a stationary orientation, and the two qubit states remain well separated even when both common-mode and differential-mode errors are applied (\fref{fig:supple_microwave_IQs}i), demonstrating the exceptional robustness of $\mathbb{S}_{Q}$ for qubit readout.  


\subsubsection{State discrimination, assignment fidelity, and resolvability} \label{sec:state_discrimination_resolvability}

\begin{figure*}[t!]
    \centering
    \includegraphics[width=0.85\linewidth]{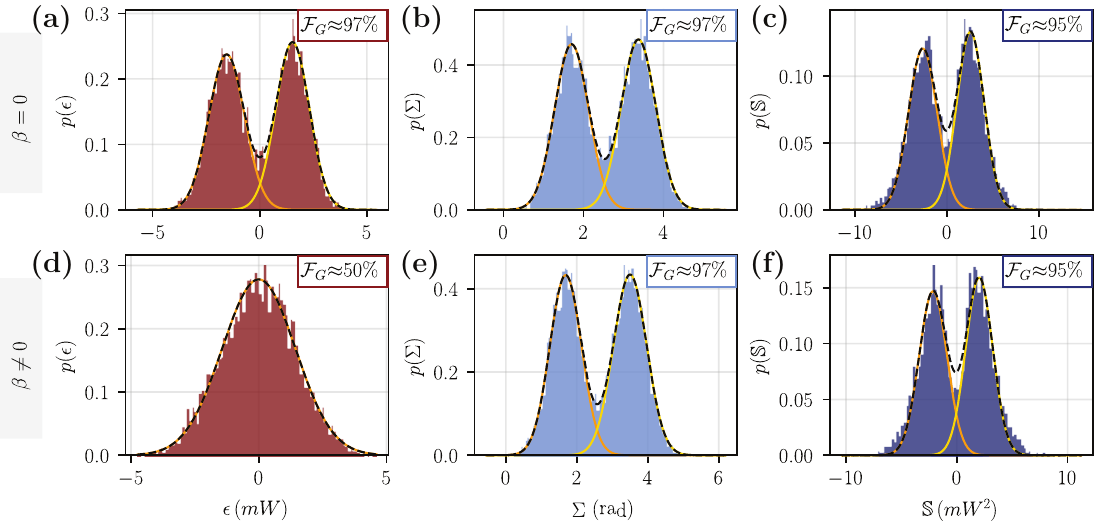}
    \caption{\textit{Superconducing-qubit readout under severe RF timing errors.} Single-shot readout probability distributions for $\epsilon_{Q}$, $\Sigma$, and $\mathbb{S}_{Q}$ constructed from the same three-tone measurements.
    (a-c) With synchronized RF electronics, all three observables yield greater than $95$\% Gaussian-fit resolvability of the two qubit states. (d-f) With deliberate RF timing errors, the conventional PDH $\epsilon_{Q}$ loses state discrimination, while $\Sigma$ and $\mathbb{S}_{Q}$ remain resolvable. The quoted $\mathcal{F}_G$ values quantify state resolvability of the displayed distributions assuming a bimodal Gaussian distribution, and do not give an independently calibrated intrinsic qubit readout fidelity. Gaussian resolvability for $\mathbb{S}$ is expected to be an underestimate of the true resolvability.
    }
    \label{fig:microwave_histograms}
\end{figure*}

The IQ distributions in \fref{fig:supple_microwave_IQs} show qualitatively how common- and differential-mode phase errors alter the geometry of the measured qubit-state distributions. We now quantify how these changes affect state resolvability for the three scalar observables $\epsilon_Q$, $\Sigma$, and $\mathbb{S}_Q$, and then examine the statistical properties of the $\mathbb{S}_Q$ distributions in greater detail. To quantify the separation of the measured state distributions, we define the resolvability
\begin{equation}
    \mathcal{F}=1-\frac{\mathcal{O}}{2},
\end{equation}
where $\mathcal{O}$ is the integrated overlap between two individually-normalized fitted components. When the two components are modeled as Gaussians, we denote the corresponding overlap by $\mathcal{O}_G$ and define the Gaussian resolvability
\begin{equation}
    \mathcal{F}_{G}=1-\frac{\mathcal{O}_{G}}{2}.
\end{equation}
With this definition, $\mathcal{F}=50\%$ for identical distributions and approaches $100\%$ as their overlap vanishes.

Under synchronized RF conditions, all three observables comparably resolve the two qubit states, as shown in \fref{fig:microwave_histograms}(a-c). We obtain $\mathcal{F}_G=97\%$ for both $\epsilon_Q$ and $\Sigma$, and $\mathcal{F}_G=95\%$ for $\mathbb{S}_Q$. Thus, in the absence of large timing errors, the scissors observables preserve the qubit-state information available in the conventional synthetic-PDH signal. The distinction appears when RF timing errors comparable to $1/\omega_m$ are deliberately introduced. Under these degraded-synchronization conditions, the conventional PDH projection collapses into an unresolved distribution with $\mathcal{F}_G=50\%$ (\fref{fig:microwave_histograms}d). In contrast, $\Sigma$ and $\mathbb{S}_Q$ remain well separated under the same measurement conditions, retaining Gaussian resolvabilities of $97\%$ and $95\%$, respectively (\fref{fig:microwave_histograms}(e-f)). This behavior is consistent with the IQ-plane response in \fref{fig:supple_microwave_IQs}: the differential-mode error rotates the PDH signal away from its fixed readout quadrature, whereas the scissors observables reject this rotation and preserve the state-dependent information contained in the three-tone measurement.

The quoted $\mathcal{F}_G$ values quantify the resolvability of the projected distributions rather than an independently calibrated fundamental qubit readout fidelity. They do not separately remove state-preparation errors, qubit decay during the measurement window or measurement-induced transitions. Notably, $\mathbb{S}_{Q}$ yields a slightly lower Gaussian resolvability ($\mathcal{F}_G=95\%$) compared to $\Sigma$ ($\mathcal{F}_G=97\%$). The slightly lower Gaussian resolvability obtained for $\mathbb{S}_Q$ does not reflect a corresponding reduction in the separation of the underlying state distributions. Because $\mathbb{S}_Q$ is constructed nonlinearly from the measured quadratures, its projected distributions are weakly asymmetric and are not perfectly described by Gaussian functions. We therefore examine the $\mathbb{S}_Q$ distributions in more detail below, distinguishing state-assignment performance from the resolvability of the measured distributions. We show that a skew-normal model provides a better empirical description of the $\mathbb{S}_{Q}$ distributions.

To perform this analysis, the two-dimensional IQ data is first projected onto the direction that maximizes the separation between the measured state centroids. For an IQ coordinate $\mathbf{r} = (I, Q)$, the states centroids are 
\[ \bar{\mathbf{r}}_g = \langle \mathbf{r}\rangle_g, \quad \bar{\mathbf{r}}_e = \langle \mathbf{r}\rangle_e,\]
and the unit vector joining them is aligned along the axis separating the two centroids, and is given by 
    \[ \hat{\mathbf{u}} = \frac{\mathbf{r}_{e} - \mathbf{r}_{g}}{\left\lVert\bar{\mathbf{r}}_e - \bar{\mathbf{r}}_{g}\right\rVert}.\]
Each single-shot IQ vector $\mathbf{r}$ is then projected relative to the midpoint of the distributions, \[\mathbf{r}_0 = \frac{\bar{\mathbf{r}}_g + \bar{\mathbf{r}}_e}{2}\]
yeilding the scalar coordinate 
\[ x = (\mathbf{r} - \mathbf{r}_0) \cdot \mathbf{u}.\]
This projected coordinate $x$ forms the basis for the histogram analysis used to benchmark readout performance and the projection axes are shown in black dashed lines in the IQ plots (\fref{fig:supple_microwave_IQs}).  

We analyze the projected $\mathbb{S}_Q$ distributions using three complementary metrics described below. The first uses the known preparation labels and measures the assignment performance. The remaining two discard those labels and quantify how well the results can be divided into two separate distributions.

\begin{figure} [t]
    \centering
    \includegraphics[width=0.8\linewidth]{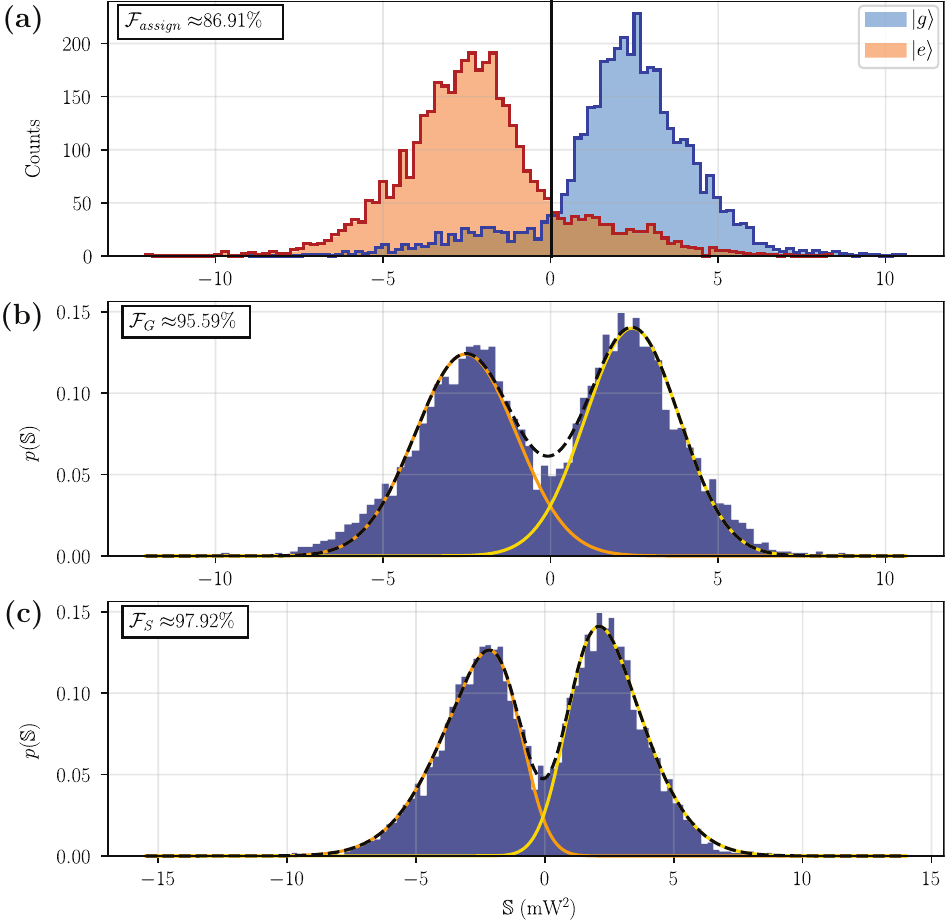}
    \caption{\textit{Assignment fidelity and resolvability of projected scissors readout.}
    Histograms obtained by projecting the single-shot scissors coordinates onto the axis joining the measured state centroids. 
    (a) Distributions conditioned on nominal preparation of the qubit states $\ket{g}$ and $\ket{e}$. The vertical black line marks the threshold that maximizes the assignment fidelity $\mathcal{F}_{assign}$. 
    (b) Histogram of the unlabeled data fit to a two-component Gaussian model. The quoted Gaussian resolvability $\mathcal{F}_{G}$ is obtained from the overlap of the two individually normalized fitted components.
    (c) The same unlabeled data as in (b) fit to a two-component skew-normal model, which better captures the asymmetric tails of the projected distributions. The corresponding skew-normal resolvability $\mathcal{F}_{S}$ is calculated using the same component-overlap definition.
    }
    \label{fig:supple_microwave_histograms}
\end{figure}

For the labeled analysis, the projected outcomes obtained after preparing $\ket{g}$ and $\ket{e}$ are histogrammed separately. For a threshold $t$, one possible assignment orientation identifies outcomes with $x\leq t$ as $\ket{g}$ and outcomes with $x > t$ as $\ket{e}$. The corresponding assignment fidelity is 
\begin{equation}
    \mathcal{F}_{g<}(t) = \frac{1}{2}\left[ \frac{N_g(x\leq t)}{N_g} + \frac{N_e(x > t)}{N_e} \right].
\end{equation}
Because the sign of the projection coordinate is arbitrary, the opposite orientation is also tested:
\begin{equation}
    \mathcal{F}_{g>}(t) = \frac{1}{2}\left[ \frac{N_g(x > t)}{N_g} + \frac{N_e(x \leq t)}{N_e} \right].
\end{equation}
The reported assignment fidelity (\fref{fig:supple_microwave_histograms}a) is 
\begin{equation}
    \mathcal{F}_{assign} = \max_{t}\{\mathcal{F}_{g<}(t), \mathcal{F}_{g>}(t)\} = 
    1 - \frac{1}{2} \left[ P(e|g) + P(g|e) \right]= 86.91\%
\end{equation}
at the optimal threshold obtained by varying $t$ across the projected data. This metric captures the macroscopic performance of the experiment, which includes events such as imperfect state preparation, qubit relaxation, thermal population, leakage, or other errors occurring before the final discrimination step. The labelled distributions are shown in \fref{fig:supple_microwave_histograms}a with the optimal threshold indicated by the vertical solid black line. 

To decouple the resolving power of the observables from the initialization errors, we instead quantify the distribution resolvability. To do this, the datasets for the two qubit states are pooled together with the labels discarded. We first assume the distributions are Gaussians and the resulting bimodal histogram is fit to the sum of two Gaussian components,
\[ y_{G}(x) = g_1 (x) + g_2 (x),\]
with 
\[ g_i (x) = A_i \exp{\left[- \frac{(x - \mu_i)^{2}}{2\sigma_i^{2}} \right]}.\]
The fitted components are individually normalized,
\[ p_i (x) = \frac{g_i (x)}{\int g_i (x')dx'},\]
and their overlap is defined as 
\begin{equation}
    \mathcal{O}_{G} = \int \min \left[p_1(x), p_2(x)\right]dx.
\end{equation}
The Gaussian resolvability is then  
\[\mathcal{F}_{G} = 1 - \frac{1}{2}\mathcal{O}_{G},\]
where $\mathcal{F}_{G} = 0.5$ (or $50\%$) for two identical distributions and $\mathcal{F}_{G} = 1$ (or $100\%$) for distributions with no overlap. This Gaussian model yields Guassian resolvability of $\mathcal{F}_{G} = 95.59\%$) as shown in \fref{fig:supple_microwave_histograms}b but should not be interpreted as a calibrated qubit-assignment fidelity because it is inferred from the histogram rather than from the preparation labels. 

Finally, because the nonlinear construction of $\mathbb{S}_Q$ produces weakly asymmetric projected distributions that are not perfectly Gaussian, we use a two-component skew-normal mixture to provide a better empirical fit. For each component, define 
\[ z_i = \frac{x - \mu_i}{\sigma_i},\]
and the normalized skew-normal probability density
\[ s_i (x) = \frac{1}{\sigma_i \sqrt{2\pi}} \exp{\left( - \frac{z_i^2}{2}\right)} \left[1 + \mathrm{erf}\left( \frac{\alpha_i z_i}{\sqrt{2}}\right) \right].\]
Here, the pooled distribution is modeled as 
\[ y_{S}(x) = w_1 s_1(x) + w_2 s_2 (x),\]
where the weights $w$ are constrained as $w_2 = 1 - w_1$. The skew parameters $\alpha_i$ control the direction and magnitude of the left/right asymmetry of the distribution. When $\alpha_i  = 0$, the corresponding component reduces to an ordinary Gaussian distribution. Using the same overlap definition as above,
\begin{equation}
    \mathcal{O}_{S} = \int\min\left[s_1(x), s_2(x)\right]dx,
\end{equation}
the skew-normal resolvability is 
\[\mathcal{F}_{S} = 1 - \frac{1}{2}\mathcal{O}_{S}.\]
The skew-normal model follows the asymmetric histogram more closely, with the fit shown in \fref{fig:supple_microwave_histograms}c giving $\mathcal{F}_{S} = 97.92\%$.

\end{document}